\input epsf.tex
\documentclass[12pt]{article}
\usepackage{psfig,picinpar,floatflt,amssymb,epsf}

\newcommand{\resection}[1]{\setcounter{equation}{0}\section{#1}}
\newcommand{\appsection}{\addtocounter{section}{1} \setcounter{equation}{0}
			 \section*{Appendix \Alph{section}}}

\newcommand{\EQ}{\begin{equation}}
\newcommand{\EN}{\end{equation}}
\newcommand{\bea}{\begin{eqnarray}}
\newcommand{\ena}{\end{eqnarray}}
\newcommand{\hs}{\hspace{0.1cm}}

\newcommand{\sn}{\rm{sn}}
\newcommand{\cn}{\rm{cn}}
\newcommand{\dn}{\rm{dn}}
\newcommand{\ka}{{\bf{K}}}

\newcommand{\be}{\beta}

\begin{document}
\setcounter{page}{0}
\topmargin 0pt
\oddsidemargin 5mm
\renewcommand{\thefootnote}{\fnsymbol{footnote}}
\newpage
\setcounter{page}{0}
\begin{titlepage}
\begin{flushright}
Bicocca--FT--99--20 
\end{flushright}
\vspace{0.5cm}
\begin{center}
{\large {\bf A Quantum Field Theory with Infinite Resonance States}}\\
\vspace{1.5cm}
{\large G. Mussardo$^{a,b}$ and S. Penati$^{c,d}$}\\
\vspace{0.5cm}
{\em $^{a}$ Dipartimento di Fisica, Universit\'a dell'Insubria, Como}\\
\vspace{0.2cm}
{\em $^{b}$Istituto Nazionale di Fisica Nucleare, Sezione di Trieste}\\
\vspace{0.2cm}
{\em $^c$ Dipartimento di Fisica, Universit\'a di Milano--Bicocca}\\
{\em via Emanueli 15, I--20126 Milano, Italy }\\
\vspace{0.2cm}
{\em $^d$ Istituto Nazionale di Fisica Nucleare, Sezione di Milano} 
\end{center}
\vspace{1.2cm}

\renewcommand{\thefootnote}{\arabic{footnote}}
\setcounter{footnote}{0}

\begin{abstract} 
\noindent 
We study an integrable quantum field theory of a single stable 
particle with an infinite number of resonance states. The exact 
$S$--matrix of the model is expressed in terms of Jacobian 
elliptic functions which encode the resonance poles inherently. 
In the limit $l \rightarrow 0$, with $l$ the modulus of the 
Jacobian elliptic function, it reduces to the Sinh--Gordon 
$S$--matrix. We address the problem of computing the Form 
Factors of the model by studying their monodromy and recursive 
equations. These equations turn out to possess infinitely many 
solutions for any given number of external particles. This 
infinite spectrum of solutions may be related to the irrational 
nature of the underlying Conformal Field Theory reached in the 
ultraviolet limit. We also discuss an elliptic version of the 
thermal massive Ising model which is obtained by a particular 
value of the coupling constant.  
\end{abstract}
\end{titlepage}

\newpage

\resection{Introduction}

The search for solvable but realistic models in theoretical 
physics has always resulted in valuable by--products. In this 
respect, the subject of two--dimensional relativistic Quantum 
Field Theories (QFT) has played a prominent role in the 
last decade, since numerous and important results have been 
achieved on non--perturbative aspects of strongly interacting 
systems. Remarkable applications to phenomena 
which occur in statistical or condensed matter systems 
where the dimensionality is effectively reduced have 
been made (see for instance \cite{ISZ,Tsvelik,GMrep}). 
In addition, the large variety of two--dimensional exactly 
solvable models -- obtained both in the scale--invariant 
regime described by Conformal Field Theories or along 
the integrable directions which depart from them -- have 
often been the theorist's ideal playground for understanding  
(at least qualitatively) phenomena which take place in 
high--energy physics. The $S$--matrix approach proposed by 
Zamolodchikov \cite{Zam} for studying the massive regime of 
integrable Renormalization Group flows has been particularly 
successful in answering important questions concerning the 
spectrum of QFT, the particle interactions and the behaviour 
of correlators of their local fields. 

The aim of this paper is to address the analysis of one of 
the most striking aspects of scattering experiments, i.e. 
the occurrence of resonances. The physical effects induced 
by resonances have been the subject of some interesting 
publications in the past. Al. Zamolodchikov, for instance, 
has shown in \cite{roaming} that a single resonance state 
can induce a remarkable pattern of roaming Renormalization 
Group trajectories which have the property to pass by very 
closely all minimal unitary models of Conformal Field Theory, 
finally ending in the massive phase of the Ising model. Another 
model, this time with an infinite number of resonance states, has 
also been considered by A. Zamolodchikov in relation with a QFT 
which is characterised by a dynamical $Z_4$ symmetry \cite{Z4Zam}: 
this is a theory with multi--channel scattering amplitudes which 
are forced by the Yang--Baxter equations to be expressed in terms 
of Jacobian elliptic functions. 

The model we propose in this paper is quite close in spirit to the 
latter one by A. Zamolodchikov \cite{Z4Zam} but has the important 
advantage of being simpler and therefore amenable for a more 
detailed analysis of its off--shell properties. More specifically, 
we will consider an integrable QFT made of a stable, self--conjugated 
scalar excitation $A$ of mass ${\cal M}$ but with an infinite 
series of resonance states which emerge as virtual unstable 
particles in the scattering amplitudes. As a function of the 
rapidity variable, its two--body $S$--matrix is expressed in 
terms of Jacobian elliptic functions, which present a periodicity 
along the real axis. This causes the appeareance of the infinite 
number of resonances. In ordinary four--dimensional QFT, resonances are responsible 
for sharp peaks observed in the total cross--section of the 
scattering processes as a function of the energy or equivalently 
in abrupt changes of the phase--shifts. In two--dimensional 
integrable QFT, on the other hand, production processes are 
forbidden and therefore the concept itself of ``total cross 
section'' is not particularly usueful. We will show, however, 
that resonances are associated to rapid jumps of the phase--shift. 

The presence of an infinite number of resonance states deeply 
affects the physical properties of the system, the analytic 
structure of the $S$--matrix and in particular the short--distance
behaviour of the correlators. This can been seen by studying 
the spectral series representations of the correlators, based 
on the Form Factors of the theory. Concerning the analysis 
of the Form Factors themselves, the main novelty consists 
in the infinite number of solutions of the Form Factors 
equations which can be found in this case. The infinite 
spectrum of operators defined by these solutions may be 
related to the irrational nature of the underlying Conformal 
Field Theory reached in the ultraviolet limit. 

The paper is organised as follows: in sect.\,2 we discuss the 
general properties of our model. Its $S$--matrix is the 
simplest example with an infinite series of resonance states. 
Since it reduces to the Sinh--Gordon $S$--matrix in the limit 
$l\rightarrow 0$ ($l$ is the module of the Jacobi elliptic 
functions), the QFT defined by such an $S$--matrix 
will be referred to as Elliptic Sinh--Gordon model (ESG). In 
sect.\,3 we briefly outline the main properties of the Form 
Factors of our relativistic integrable QFT and discuss the new 
features which arise in a theory with double periodicity. 
Section 4 is devoted to the analysis of the Form Factors of the  
Elliptic Ising Model, i.e. a model with infinite resonance states 
which reduces to the usual Ising model in the limit $l \rightarrow 0$. 
This model may be obtained by a particular analytic continuation 
of the coupling constant of the ESG model. In sect.\,5 we address the
computation of the Form Factors of the Elliptic Sinh--Gordon model: 
we present the computation of the minimal Form Factor $F(\beta)$ and 
investigate its analytic structure. By using the functional equation 
satisfied by $F(\beta)$, we derive the recursive equations of the 
Form Factors and obtain their first solutions. The main result 
of this section is that the solutions of the FF equations for 
a given number of external particles span an infinite dimensional 
vector space. In sect.\,6 we present our conclusions. The paper also
contains three appendices: appendix A presents useful mathematical 
identities used in the text, appendix B contains the derivation of 
the Fourier series of the $S$--matrix and finally appendix C presents 
the calculation of the minimal Form Factor of the Elliptic 
Sinh--Gordon model. 

\resection{Elastic $S$-matrix with Resonance States}

We consider a two--dimensional integrable QFT describing 
a stable, self--conjugated particle $A$ of mass ${\cal M}$. 
This theory is assumed to be invariant under a $Z_2$ symmetry  
realised by $A \rightarrow - A$. Its on--shell properties 
are encoded into its elastic $S$--matrix. In virtue of 
integrability, the $S$--matrix satisfies the factorization 
condition \cite{ZZ}, therefore it is sufficient to focalise only 
on the two--body elastic scattering process. Let $s =(p_1+p_2)^2$ 
be the Mandelstam variable of the scattering process and $S(s)$ the 
two--body elastic scattering amplitude of the particle $A$. 
The function $S(s)$ is usually expected to have two elastic 
branch--cuts along the real axis at the two--particle 
thresholds $ s \leq 0$ and $s \geq 4 {\cal M}^2$. 
These branch--cuts can be unfolded as follows. Let $\beta$ 
be the rapidity variable which parameterises the relativistic 
dispersion relations
\EQ
E = {\cal M}\cosh\beta \,\,\,\,\,\,\, ; \,\,\,\,\,\,\,\, 
p = {\cal M} \sinh\beta 
\,\,\,.
\label{dispersion}
\EN
The Mandelstam variables $s$ and $t$ of the two--body 
scattering processes are then expressed by 
\EQ
\begin{array}{l}
s = (p_1 + p_2)^2 = 2 {\cal M}^2 \left[1+ \cosh\beta_{12} \right]
\,\,\,;\\
t = (p_1 - p_2)^2 = 2 {\cal M}^2 \left[1+ \cosh(i\pi-\beta_{12})\right]
\,\,\,.
\end{array}
\EN 
($\beta_{12} = \beta_1-\beta_2$). Hence the upper and lower edges 
of the branch--cut which starts at $s = 4 {\cal M}^2$ are mapped 
onto the positive and negative real semi--axes of $\beta$ 
respectively. The upper and lower edges of the other branch--cut 
are conversely mapped onto the positive and negative values of 
$\beta$ along the line ${\rm Im} \beta = i \pi$. The $S$--matrix  
$S(\beta_{12})$ becomes a meromorphic function in the 
$\beta_{12}$--plane. It satisfies the unitarity and crossing 
symmetry conditions
\EQ
\begin{array}{l}
S(\beta) S(-\beta) = 1 \,\,\,;\\
S(i \pi - \beta) = S(\beta) \,\,\,.
\end{array}
\label{unitcro}
\EN 
These equations automatically imply that $S(\beta)$ is a 
periodic function along the imaginary axis of the rapidity 
variable, i.e.  
\EQ
S(\beta + 2 \pi i) = S(\beta) \,\,\,.
\label{periodim}
\EN 
This periodicity simply expresses the double--sheet structure 
of the Riemann surface with respect to the Mandelstam variable 
$s$ where the scattering amplitude is meromorphic. The physical 
sheet can be taken then as the strip $ 0 \leq {\rm Im}\beta \leq 
\pi$ and the unphysical one as the strip $ -i\pi \leq {\rm Im} 
\beta \leq 0$. 

Let us now assume that $S(\beta)$ also presents a periodic 
behaviour along the real axis of the rapidity variable 
$\beta$ with a period $T$, i.e. 
\EQ
S(\beta) = S(\beta + T) \,\,\,.
\label{period}
\EN 
This equation, combined with the analyticity properties of 
$S(\beta)$, has far--reaching consequences. In fact, the 
simultaneous validity of the two periodicity conditions 
(\ref{periodim}) and (\ref{period}) forces $S(\beta)$ to 
be an elliptic function. Accordingly, the plane of the 
rapidity variable $\beta$ becomes tiled in terms of 
the periodic cells of the function $S(\beta)$ (Figure 1.a). 
On the plane of the Mandelstam variable $s$ the $S$--matrix
is not periodic. However, the existence of a double periodicity 
on the $\beta$--plane induces an infinite swapping between the 
two edges of the elastic branch--cuts on the $s$--plane.
The swaps are located at the infinite set of values 
$s_n = 2{\cal M}^2(1+\cosh (nT/2))$ and 
$s_n = 2{\cal M}^2(1+\cosh (i\pi -nT/2))$ for the $(4{\cal M}^2,
+\infty)$ and $(-\infty,0)$ branch--cuts, respectively. 
In fact, take $(-\frac{T}{2},\frac{T}{2})$ as fundamental 
interval of the real periodicity in $\beta$ and consider,
for instance, the branch--cut $(4M^2, +\infty)$ in the variable $s$. 
At the initial location $s= 4M^2$ ($\beta =0$) the two edges join 
together. Starting from $\beta =0$ and moving onto the negative 
values of $\beta$, we proceed along the lower edge of the branch 
cut, as far as $\beta > -\frac{T}{2}$. When this value is crossed 
and $\beta$ reaches the region $(-T,-\frac{T}{2})$, due to the 
identification $\beta \equiv \beta + T$, we are brought back to 
positive values of $\beta$, $0 < \beta < \frac{T}{2}$. On the 
other hand, this interval for $\beta$ parameterises the upper 
edge of the branch cut. Therefore, at $\beta = -\frac{T}{2}$ a 
first swap of the two edges of the branch cut has occured. 
If now $\beta$ decreases, we keep moving along the upper edge
as far as $\beta$ is greater than zero. When the rapidity crosses
this value (now identified with $\beta = T$) a second swap occurs 
and the original configuration of the two edges is restored. The 
pattern described in details for $0 < \beta < T$, indeed reproduces
itself with period $T$ in the rapidity variable. In particular,  
since the point $\beta=0$ is identified with an infinite sequence 
of points $\beta_n= nT$, the two branches return to the initial 
joined configuration an infinite number of times. For the same 
reason, the swap at $\beta = -\frac{T}{2}$ is reproduced at every 
value $\beta_{2n+1} = (2n+1)\frac{T}{2}$. The situation is completly 
analogous for the branch--cut of the $t$--channel which originally 
lies on the real negative semi--axis of the $s$ variable. Hence, 
altogether in the plane of the Mandelstam variable $s$ we have 
the analytic structure of the $S$--matrix shown in Figure 1.b.  

If $S(\beta)$ is an elliptic function, it must have poles 
and zeros, unless it is a constant. However, the existence
of poles should not spoil the causality properties 
of the scattering theory. From a mathematical point of view, this 
is equivalent to requiring a further condition on the analytic 
structure of $S(\beta)$, namely that this function should not have 
poles with a real part within the physical sheet. 

It is now easy to determine the simplest $S$--matrix which 
satisfies all of the above constraints, i.e. the unitarity and 
crossing equations (\ref{unitcro}), the periodicity equation  
(\ref{period}) and the causality condition. It may be written 
as 
\EQ
S(\beta,a) = 
\frac
{\sn\left(\frac{\ka (\beta - i \pi {\it a})}{i \pi}\right)}
{\sn\left(\frac{\ka (\beta + i \pi {\it a})}{i \pi}\right)}
\frac
{\cn\left(\frac{\ka (\beta + i \pi {\it a})}{i \pi}\right)}
{\cn\left(\frac{\ka (\beta - i \pi {\it a})}{i \pi}\right)}
\frac
{\dn\left(\frac{\ka (\beta + i \pi {\it a})}{i \pi}\right)}
{\dn\left(\frac{\ka (\beta - i \pi {\it a})}{i \pi}\right)} 
\,\,\,,
\label{ellipS}
\EN 
where $\sn({\it x})$, $\cn({\it x})$ and $\dn({\it x})$ 
are the Jacobian elliptic 
functions and $\ka$ is the complete elliptic integral, both 
of modulus $l$ (see Appendix A for a review of the main 
properties of the Jacobian elliptic functions and some of their
useful identities). By using the addition theorems (\ref{addition}), 
the above function can be equivalently expressed as 
\EQ
S(\beta,a) = \frac{
\sn\left(\frac{2 \ka i \beta}{\pi}\right) + \sn(2 \ka {\it a})}
{\sn\left(\frac{2 \ka i \beta}{\pi}\right) - \sn(2 \ka {\it a})} 
\,\,\,.
\label{2ellipS}
\EN 
Exploiting some of the identities listed in Appendix A, it is
easy to check the validity of eqs.\,(\ref{unitcro}).
Moreover, the $S$--matrix satisfies the real 
periodicity condition (\ref{period}) with period
\EQ              
T = \pi\, \frac{\ka'}{\ka} \,\,\,,
\label{expperiod}
\EN 
where $\ka'$ is the complete elliptic integral of the 
the complementary modulus $l'=(1-l^2)^{1/2}$.  
The parameter $a$ entering the expression (\ref{ellipS})
may be regarded as the coupling constant of the model. It  
must be selected so as not to introduce poles in the 
$S$--matrix which have a real part in the physical strip. 
As we will see below (eq. (\ref{smatrixpoles})), 
this is ensured by taking $a$ to be a real positive 
number in the interval $[0,1]$. As a matter of fact, 
this interval can be further reduced in virtue of 
an additional property of the $S$--matrix (\ref{2ellipS}).  
In fact, the $S$--matrix (\ref{2ellipS}) is invariant 
under the replacement $a \rightarrow 1 - a$,  
\EQ
S(\beta,a) = S(\beta,1-a) \,\,\, , 
\label{duality}
\EN 
so that $a$ can be restricted to the interval $[0,\frac{1}{2}]$. 
For $a = 0$ we have $S =1$, i.e. a free theory, whereas the point 
$a=\frac{1}{2}$ may be regarded as the self--dual 
point of this scattering theory. In the following, unless 
explicitly stated, $a$ will be always taken as $0 \leq a 
\leq \frac{1}{2}$. 

Notice that by taking the limit $l \rightarrow 0$, 
the real period $T$ goes to infinity and 
the $S$--matrix (\ref{2ellipS}) reduces to the one of the 
Sinh--Gordon model \cite{AFZ}
\EQ
S_{\rm ShG}(\beta,a) = \frac{\sinh\beta - i \sin\pi a}
{\sinh\beta + i \sin\pi a} \,\,\,.
\label{ShS}
\EN 
For this reason, the QFT defined by the $S$--matrix 
(\ref{2ellipS}) may be simply referred to as the 
Elliptic Sinh--Gordon Model (ESG). 

Let us consider the analytic structure of the $S$--matrix  
(\ref{ellipS}), i.e. the pattern of its poles and zeros.
Poles are either those of the numerator or come from 
the zeros of the denominator in (\ref{ellipS}). Conversely, 
zeros are either those of the numerator or come from the 
poles of the denominator. Using the results listed in 
Table 1 and taking into account possible cancellations of 
poles against zeros, we are left with the following 
infinite set of simple poles and simple zeros
\EQ
\hspace{-5mm}
\begin{array}{ll}
{\rm Poles:}~~~~
&\beta_{m,n} = ~ -i \pi a + 2 m \pi i + n T \,\,\,;\\
&\beta_{m,n} = ~ i \pi a + (2 m + 1) \pi i + n  T
\label{smatrixpoles}
\end{array}
\EN 
\EQ
\begin{array}{ll}
{\rm Zeros:}~~~~
&\beta_{m,n} = ~ i \pi a + 2 m \pi i +  n T \,\,\,;\\
&\beta_{m,n} = ~ -i \pi a + (2 m+1)  \pi i + n T
\label{smatrixzeros}
\end{array}
\EN 
where $m,n \in {\bf Z}$.   
  
By choosing in the $\beta$ plane the rectangle with vertices 
$(i\pi -\frac{T}{2},i\pi + \frac{T}{2}, -i \pi  -\frac{T}{2},
-i\pi +\frac{T}{2})$ as a fundamental domain, the analytic 
structure of the $S$--matrix is shown in Figure 2, with two 
zeros and two poles along the imaginary axis and then repeated 
on the other cells of the $\beta$--plane by periodicity. Hence, 
the simplest $S$--matrix with double periodicity is realised 
by an elliptic function of order two. With the above choice 
of the values of $a$, the poles of the scattering amplitude 
which have a real part in the $\beta$--plane all lie on the 
unphysical sheet. Instead in the physical sheet there are 
only zeros. The poles on the unphysical sheet with a real 
part correspond to a series of resonance states ${\cal R}_n$ 
with masses and decay widths given respectively by  
\EQ
\begin{array}{ll}
M_{\rm res}(n) = & 2 {\cal M} \cosh\left(n \,\frac{T}{2}\right) 
\,\cos\left(\frac{\pi a}{2}\right) \,\,\,;\\
\Gamma_{\rm res}(n) = & 2 {\cal M} \left| \sinh\left(n \,\frac{T}{2}
\right) \, \sin\left(\frac{\pi a}{2}\right) \right| \,\,\,.
\end{array}
\label{resonances}
\EN 
In the limit when the modulus $l$ of the Jacobian elliptic 
function goes to zero (i.e. $T \rightarrow \infty$), the mass 
of all resonances becomes infinitely heavy and therefore these 
states completly decouple from the theory. 
 
It is also useful to express the S--matrix (\ref{ellipS}) 
in an exponential form, a representation which will result  
particularly convenient for the computation of the Form 
Factors of the theory presented in the next sections. As 
shown in details in Appendix B, in the strip $0\leq |{\rm Im} 
\beta| < \pi a$ we have the following equivalent representation 
of the $S$--matrix of the ESG model
\EQ 
S(\beta) = - \exp\left[2 i \,
\sum_{n=1}^{+\infty} \frac{1}{n} \, \left(
\frac{\cosh\left(\frac{n
\pi^2 (1 - 2 a)}{T}\right)} {\cosh\left(\frac{n \pi^2}{T}\right)} 
- (-1)^n \right) 
\, \sin\left(\frac{2 n \pi}{T} \beta\right)
\right] \,\,\, . 
\label{firstexp} 
\EN 
If we now use the Fourier series identity (\ref{usefulidentity}) 
in order to express the factor $-1$ in front of 
(\ref{firstexp}) as 
\EQ 
-1 = \exp\left[-4 i
\,\sum_{n=1}^{+\infty} \frac{1}{2 n -1} 
\sin\left(\frac{2\pi (2 n - 1) \beta}{T}\right) 
\right] \,\,\,, 
\label{Ising} 
\EN 
the expression (\ref{firstexp}) can be equivalently written as  
\EQ 
S(\beta) =
\exp\left[-4 i \, \sum_{n=1}^{+\infty} \frac{1}{n} \, \frac{
\sinh\left(\frac{n a \pi^2}{T}\right) \, \sinh\left(\frac{n (1-a)
\pi^2}{T}\right) } {\cosh\left(\frac{n \pi^2}{T}\right)} \,
\sin\left(\frac{2 n \pi}{T} \beta\right)\right] \,\,\,. 
\label{secondexp}
\EN
In the limit $T \rightarrow \infty$, the sum in (\ref{secondexp}) 
can be converted into an integral and therefore one recovers the 
exponential representation of the Sinh--Gordon  $S$--matrix \cite{FMS} 
\EQ
S_{\rm ShG}(\beta,a) = \exp\,\left[-4 i \,
\int_0^{\infty} \frac{dx}{x}
\frac{\sinh\left(\frac{x a}{2}\right)
\sinh\left(\frac{x}{2}(1-a)\right)}{\cosh \frac{x}{2}}
\sin\left(\frac{x \beta}{\pi}\right)\right] \,\,\,.
\label{integralShG}
\EN

\vskip 15pt

By using the expression (\ref{secondexp}) we can read 
the phase--shift $\delta(\beta)$ of the model defined by
\EQ
S(\beta) = e^{2i\delta(\beta)} \,\,\,. 
\EN
The phase--shift is obviously a quantity which is defined 
modulus multiplies of $\pi$. Therefore, by adding to 
$\delta(\beta)$ a convenient multiplies of $\pi$ we 
can always consider $\delta(\beta)$ to be a monotonic 
function of $\beta$. When $\beta$ approaches the locations  
of the resonances ${\rm Re}\beta_n = n T$, the  phase--shift
$\delta(\beta)$ sharply increases its value by $\pi$ (see Figure 
3). 

\vskip 30pt

A particularly simple but fascinating example of integrable
theory with an infinite number of resonances is given by the 
$S$--matrix $S = -1$, when expressed in the form (\ref{Ising}). 
The phase--shift of such a model can be taken as the staircase 
function with values given by all multiplies of $\frac{\pi}{2}$ 
and jumps localised in correspondence to the points $\beta_n = 
n T/2$ (Figure 4). A way to understand this unexpectedly rich 
structure hidden in $S=-1$ is to notice that the scattering 
theory described by (\ref{Ising}) can be obtained as a particular 
analytic continuation of the elliptic expression (\ref{2ellipS}). 
Indeed, the $S$--matrix (\ref{2ellipS}) approaches the value $-1$
every time the parameter $a$ takes one of the infinite values 
\EQ
a_c = m + (2n+1) i \,\frac{T}{2\pi}   \,\quad ,\qquad m \in {\bf N}, \, 
\, n \in {\bf Z}
\EN
corresponding to the locations of the poles of $\sn(2\ka {\it a})$.
At these critical values of $a$, the infinite number of poles 
and zeros of (\ref{2ellipS}) cancel each other over the entire  
$\beta$--plane, giving rise to the constant function $S = -1$. 
In Figure 5 this pattern is described in details for the simplest 
choice $a_c = -i \frac{T}{2\pi}$. In this case, for a theory 
defined at an infinitesimal displacement $a = a_c + \delta$ 
from the critical value $a_c$, the location of poles and zeros is
\EQ
\hspace{-20mm}
\begin{array}{ll}
{\rm Poles:}~~~~
&\beta_{m,n}^{(-)} = ~ (2n-1)\frac{T}{2} + 2 m \pi i - 
i\pi \delta \,\,\,;\\
&\beta_{m,n}^{(+)} = ~ (2n+1)\frac{T}{2} + (2 m + 1) \pi i +
i\pi \delta \,\,\, ,
\label{criticalpoles}
\end{array}
\EN 
\EQ
\begin{array}{ll}
{\rm Zeros:}~~~~
&\beta_{m,n}^{(+)} = ~ (2n+1)\frac{T}{2} + 2 m \pi i + i\pi 
\delta \,\,\,;\\
&\beta_{m,n}^{(-)} = ~ (2n-1)\frac{T}{2} + (2 m+1)  \pi i 
-i \pi \delta \,\,\, ,
\label{criticalzeros}
\end{array}
\EN 
where $m,n \in {\bf Z}$. As shown in Figure 5, in the limit $\delta
\rightarrow 0$ cancellations occur between $\beta_{m,n+1}^{(-)}$, 
$\beta_{m,n}^{(+)}$ poles and $\beta_{m,n}^{(+)}$, $\beta_{m,n+1}^{(-)}$
zeros. The $S$--matrix $S=-1$ defined in this limit inherits all the 
analytic structure of the ESG model. In particular, the infinite 
number of odd values $\beta_{2n+1} = (2n+1)T/2$ corresponding to 
rapid increasings of the phase--shift can be interpreted as 
associated to resonance states coming from the theory (\ref{2ellipS}) 
defined at $a= a_c + \delta $. Near the values $\beta = i\pi + 
\beta_{2n+1}$ the $S$--matrix becomes 
\EQ
S(\beta) \,\simeq\,  - \frac{ i \pi \delta }
{\beta - i\pi - \beta_{2n+1}} \,\,\, ,
\label{residueresonance}
\EN 
which corresponds to the diagram in Figure 6. In general, at the 
poles $\beta_{ij} = i u_{ij}^k $ corresponding to a bound state 
$A_k$ produced by the fusion of two particles $A_i$ and $A_j$, 
the $S$--matrix satisfies
\EQ
-i\,\lim_{\beta\rightarrow i u_{ij}^k} 
(\beta-i u_{ij}^k) \,S(\beta)\,=\, \left(\Gamma_{ij}^k\right)^2 \,\,\,,
\label{bound}
\EN
where $\Gamma_{ij}^k$ is the three--particle vertex on mass--shell. 
Therefore, comparing eqs. (\ref{residueresonance}) and (\ref{bound}), 
we can interpret the resonances $R_n$ as ``bound states'' of the 
particle $A$ but on the second sheet, with the three--particle vertex 
on mass--shell given by $\Gamma_{AA}^{R_n} = i {\sqrt{\pi \delta}}$. 
The imaginary value of the three--particle vertex indicates that 
the QFT presents some non--unitary features. As bound states of the 
$Z_2$ odd particle $A$, the resonance states $R_n$ are even under 
the $Z_2$ symmetry of the model. Near the values $\beta = \beta_{2n+1}$ 
the $S$--matrix behaves as in eq.\,(\ref{residueresonance}) but with an
opposite sign for the residue. These poles correspond to resonance 
states in the crossed channel.   

The even values $\beta_{2n} = nT$ corresponding to the other rapid
increasings of the phase --shift have, on the other hand, a 
different physical interpretation. They do not come from genuine 
bound state resonances but simply correspond to ``kinematic'' 
resonances associated to the infinite replicae of the initial
joint--point of the elastic branch--cuts on the $s$--plane 
(see Figure 1.b).  

The QFT theory described by the $S$--matrix (\ref{Ising}) 
will be referred to as the Elliptic Ising model (EIM).
The detailed study of its operator content will be 
discussed in Section 4. 

\resection{Form Factors in Integrable QFT}

The on--shell dynamics of a system is encoded into its $S$--matrix. 
In order to go off--shell and compute the correlation functions 
of the model, one can use their spectral series representation. 
This means that we need to compute the matrix elements of the 
quantum fields between asymptotic states. It is sufficient to 
consider the following matrix elements, known as Form Factors 
(FF) \cite{Karowski,Smirnov}
\EQ
F_n^{\cal O} (\beta_1,\beta_2,\ldots,\beta_n) \,=\,
\langle 0\mid{\cal O}(0,0)\mid \beta_1,\beta_2,\ldots,\beta_n\rangle_{in}
\,\,\, . 
\label{FF}
\EN
In this section we will briefly outline the main properties of 
the FF in a generic two--dimensional integrable QFT and we will 
comment on some features which arise in the case of an 
$S$--matrix with a real period. A more detailed discussion of 
these features is postponed to the next sections. For simplicity 
we only consider the FF of scalar and hermitian operators ${\cal O}$. 

The FF of a QFT are known to be severely constrained by the 
relativistic symmetries. First of all, they only depend on 
the difference of the rapidities $\beta_{ij}$ 
\EQ
F_n^{\cal O}(\beta_1,\beta_2,\ldots,\beta_n) \,=\,
F_n^{\cal O}(\beta_{12},\beta_{13},\ldots,\beta_{ij},\ldots) 
\,\,\,\,, i<j
\,\,\, .
\EN

Except for pole singularities due to the one-particle intermediate 
states in all sub-channels, the form factors $F_n$ are expected to 
be analytic functions inside the physical strip  $0 < {\rm Im } 
\be_{ij} < \pi$. Moreover, they satisfy the monodromy equations 
\cite{Karowski, Smirnov}
\begin{eqnarray}
F_n^{\cal O}(\be_1,..,\be_i, \be_{i+1},.., \be_n) &=& 
F_n^{\cal O}(\be_1,..,\be_{i+1}
,\be_i ,.., \be_n) S(\beta_i-\beta_{i+1}) \,\,; 
\label{permu1}\\
F_n^{\cal O}(\be_1+2 \pi i,.., \be_{n-1}, \be_n ) &=& e^{2\pi i 
\omega} \, 
F_n^{\cal O}(\be_2 ,\ldots,\be_n, 
\be_1) \,\,\,.
\nonumber 
\end{eqnarray}
The parameter $\omega$ is the index of mutual locality of the
operator  ${\cal O}$ with respect to the particle state $A$ 
($\omega =0, 1/2 $ for local and semi--local operators, 
respectively).

In the simplest case $n=2$, the above equations become 
\EQ
\begin{array}{ccl}
F_2^{\cal O}(\beta)&=&F_2^{\cal O}(-\beta)S(\beta) \,\, ,\\
F_2^{\cal O}(i\pi-\beta)&=&e^{2\pi i \omega}\, F_2^{\cal O}(i\pi+\beta)
\,\,\, . \end{array} 
\label{F2}
\EN
By exploiting the factorization properties of integrable QFT, the 
general solution of (\ref{permu1}) can be found by making the ansatz 
\EQ
F_n^{\cal O}(\beta_1,\dots,\beta_n) = K_n(\beta_1,\dots,\beta_n) 
\,\prod_{i<j}F_{\rm min}
(\beta_{ij})  \,\, , 
\label{parametrization}
\EN
where $F_{\rm min}(\beta)$ is a solution of eqs.\,(\ref{F2}), 
analytic in $0 < $ Im $\beta < \pi$. The remaining factors 
$K_n$ therefore satisfy the monodromy equations
with $S=1$, i.e. they are completely symmetric, $2 
\pi i$-periodic functions of the $\beta_{i}$. Moreover, they 
contain all the physical poles expected in the Form Factor 
under investigation.  

In the case of a period $S$--matrix, the general solutions $F_n$ of 
the monodromy equations
are not a--priori constrained to have any real periodicity in the 
rapidity variables. This can be easily seen from their parameterization
(\ref{parametrization}); the monodromy equations with $S=1$ which 
determine the functions $K_n$ do not contain any informations
about the periodic nature of the dynamics under consideration. 
However, for the two--particle FF the solutions of eqs.\,(\ref{F2}) 
may be chosen to satisfy the extra condition 
\EQ
F_2^{\cal O}(\beta + T) = 
\pm F_2^{\cal O}(\beta) 
\,\,\,.
\label{periodicityFF}
\EN 
In order to prove it, consider the first of eqs.\,(\ref{F2}) and 
make the substitution $\beta \rightarrow \beta + T$. Since 
$S(\beta)$ does not change under this shift in $\beta$, if 
$F_2(\beta)$ is a solution of eqs.\,(\ref{F2}), then $F_2(\beta+T)$ 
is also a solution 
\EQ
F_2^{\cal O}(\beta + T) = 
S(\beta) F_2^{\cal O}(-\beta - T) \,\,\,.
\label{shift}
\EN 
By taking the ratio of eq.\,(\ref{shift})
and the first of (\ref{F2}) we have 
\EQ
\frac{F_2^{\cal O}(\beta + T)}{F_2^{\cal O}(\beta)} = 
\frac{F_2^{\cal O}(-\beta-T)}{F_2^{\cal O}(-\beta)} 
\,\,\,.
\label{ratio}
\EN 
Therefore, if we assume $F^{\cal O}(\beta + T)$ to be 
proportional to $F^{\cal O}(\beta)$, i.e. $F^{\cal O}(\beta+T) = 
a F^{\cal O}(\beta)$, then eq.\,(\ref{ratio}) implies 
$a^2=1$, i.e. $a = \pm 1$. In particular, for $a=-1$ there is a
doubling of the real periodicity in the two--particle Form Factor with
respect to the $S$--matrix.  

It is simple to obtain an explicit $T$--periodic solution of 
eqs.\,(\ref{F2}) in 
the case of a local operator ($\omega =0$) when the $T$--periodic 
$S$--matrix is expressed in the form 
\EQ
S(\beta) = 
\exp\left[-i \sum_{n=1}^{+\infty} \frac{a_n}{n} 
\sin\left(\frac{2 n \pi}{T} \beta\right)\right] \,\,\,.
\label{fourierexp}
\EN 
Denoting such a solution as $F_{\rm min}(\beta)$, we have in fact 
\EQ
F_{\rm min}(\beta) = {\cal N}\,\exp\left[\sum_{n=1}^{+\infty} 
\frac{a_n}{n} 
\frac{1}{\sinh\left(\frac{2 n \pi^2}{T}\right)} 
\sin^2\left(\frac{n\pi}{T} \hat\beta\right)\right] 
\,\,\,, 
\label{fourierFFexp}
\EN 
where $\hat\beta \equiv i\pi-\beta$ and ${\cal N}$ a normalization 
factor. 

Let us consider now the pole singularities of the FF. They give 
rise to a set of recursive equations for the $F_n$ which may be 
particularly important for their explicit determination. As 
functions of the rapidity differences $\beta_{ij}$, the FF 
present in general two kinds of simple poles. The first family of 
singularities come from kinematical poles located at 
$\beta_{ij}=i\pi$. They correspond to the one-particle 
intermediate state in a subchannel of three-particle states 
which, in turn, is related to a crossing process of the elastic
$S$-matrix. The corresponding residues give rise to a recursive 
equation between the $n$-particle and the $(n+2)$--particle 
Form Factors \cite{Smirnov} 
\bea
& -i\lim_{\tilde\beta \rightarrow \beta}
(\tilde\beta - \beta)
F_{n+2}^{\cal O}(\tilde\beta+i\pi,\beta,\beta_1,\beta_2,\ldots,\beta_n)=
\nonumber \\
&~~~~~~~~~~~
\left(1 - e^{2\pi i \omega} \prod_{i=1}^n S(\beta-\beta_i)\right)\,
F_n^{\cal O}(\beta_1,\ldots,\beta_n) \,\,\, . 
\label{recursive}
\ena
For a theory with a real periodic $S$--matrix we still expect 
the appearance of kinematic poles located at $\beta_{ij}=i\pi$ 
with residue given by the above equation, whose physical origin 
is the same as in the non--periodic case (see, for instance, the 
discussion in \cite{Yurov}). However, in this case the presence 
of a periodic $S$--matrix on the r.h.s. of eq.\,(\ref{recursive})
gives additional constraints on the behaviour of the FF near the 
pole. Precisely, if we perform the shift $\beta_i \rightarrow 
\beta_i +T$, $i=1, \cdots , n$ in eq.\,(\ref{recursive}), the 
r.h.s. is left invariant due to the $T$--periodicity of $S$ and 
the traslation invariance of $F_n$ (the same conclusion is also 
obtained by shifting $\tilde\beta \rightarrow \tilde\beta +T$ and 
$\beta \rightarrow \beta +T$ on the l.h.s. of (\ref{recursive})). 
Hence, in this case the residue of $F_{n+2}^{\cal O}$ in 
$(\tilde{\beta} -\beta) = i\pi$ has to be a $T$--periodic function of 
the $n+1$ rapidities $\beta, \beta_1, \cdots ,\beta_n$. Although 
a sufficient condition to implement this requirement would be  
$F_n(\beta_1,...,\beta_n)$ periodic in all its variables, this is 
not a necessary condition as we will show in Sect. 5
by providing explicit examples.

In $F_n$ there may also be another family of poles related to the 
presence of bound states. These poles are located at the values of 
$\beta_{ij}$ where two particles fuse into a third one. Let $\beta_{ij}
=i u_{ij}^k$ be one of such poles associated to the bound state $A_k$ 
in the channel $A_i\times A_j$ and $\Gamma_{ij}^k$ the three-particle 
vertex on mass-shell. In this case, the residue equation for the 
$S$--matrix is given in (\ref{bound}) and, correspondingly, for $F_n$ 
we obtain \cite{Smirnov}  
\EQ
-i\lim_{\epsilon\rightarrow 0} \epsilon\,
F_{n+1}^{\cal O}(\beta+i \overline u_{ik}^j+\frac{\epsilon}{2},
\beta-i \overline u_{jk}^i-\frac{\epsilon}{2},\beta_1,\ldots,\beta_{n-1})
\,=\,\Gamma_{ij}^k\,F_{n}^{\cal O}(\beta,\beta_1,\ldots,\beta_{n-1}) 
\,\,\, ,
\label{respole}
\EN
where $\overline u_{ab}^c\equiv (\pi-u_{ab}^c)$. As discussed in 
sect.2, in the periodic case we cannot have genuine bound states. 
In fact, a bound state pole in the $S$--matrix at $\beta_{ij} = 
u_{ij}^k$ would be associated to an infinite chain of poles at 
$\beta_{ij} = u_{ij}^k + nT$, $n \in {\bf Z}$ with a consequent violation 
of causality. However, by considering the resonances as ``bound 
states'' on the second sheet, the above equation may be nevertheless 
used to extract some useful informations. 

In general, the FF of real periodic models are then expected to be 
ruled only by the recursive equations coming from the kinematical 
singularities, eq.\,({\ref{recursive}). In the particular case of 
the ESG model, this implies the complete decoupling between the two 
chains of FF for even or odd number of particles. This reflects 
the $Z_2$ symmetry of the model which can be used to 
classify its operators. 

Finally, we remind that in the usual non--periodic scattering 
theories the FF may be further restricted by extra conditions 
related to their asymptotic behaviour \cite{Smirnov,DM} and their 
cluster properties \cite{KM,cluster1,cluster2}. These conditions 
are often sufficient to determine uniquely the solutions of the
recursive equations for each choice of the operator ${\cal O}$.  
In the periodic case, as it will be clear later, we deal instead 
with periodic functions for which the lack of asymptotic extra 
conditions leads eventually to an anavoidable arbitrariness in 
the determination of the FF.

\resection{Form Factors of Ising Model Realizations}

The QFT defined by the $S$--matrix $S = -1$ is usually 
identified with the thermal deformation of the critical 
Ising model with central charge $C=\frac{1}{2}$ \cite{Koberle}. 
In fact, the solution of the FF equations in the space of 
hyperbolic functions permits the full reconstruction of all 
correlators of this statistical model and a precise check 
of its ultraviolet limit \cite{Karowski,Yurov,CMform}. In 
order to point out the important differences coming from 
an elliptic interpretation of this scattering theory, it 
is worth summarising first the main results for the correlation 
functions of the standard thermal Ising model. 

\subsection{The Standard Thermal Ising Model} 

With $S = -1$, the simplest solutions of eqs.\,(\ref{F2}) 
in the space of hyperbolic functions are given by 
\EQ
F_{\rm min}(\beta) = \left\{
\begin{array}{cc}
f(\beta) = \sinh\frac{\beta}{2}\,\,\, , 
& \,\,\,\mbox{if $\omega =0$} \,\,\,
;\\ {\cal F}(\beta) = \tanh\frac{\beta}{2} \,\,\, ,
&\,\,\, 
\mbox{if $\omega =\frac{1}{2}$} 
\,\,\,.
\end{array}
\right. 
\label{FminIsing}
\EN 
The kinematical recursive equations for the FF are particularly 
simple in this theory 
\EQ
-i\lim_{\tilde\beta \rightarrow \beta}
(\tilde\beta - \beta)
F_{n+2}^{\cal O}(\tilde\beta+i\pi,\beta,\beta_1,\beta_2,\ldots,\beta_n)=
\left(1- e^{2\pi i \omega}\,(-1)^n \right)\,
F_n(\beta_1,\ldots,\beta_n) \,\,\, . 
\label{recursiveIsing}
\EN
Hence, the FF of local operators ($\omega =0$) do not have 
kinematical poles when $n$ is an even integer whereas their 
residue is independent of the number of external particles 
when $n$ is an odd integer. The situation is reversed for 
the FF of semi--local operators ($\omega = \frac{1}{2}$). 

All fields are naturally divided into two classes. The first 
consists of $Z_2$ even, local or $Z_2$ odd, semi--local fields. 
Local operators of this class have non--zero FF only between 
states with a finite number of external particles. The first 
representative of this class is given by the trace of the 
stress--energy tensor $\Theta(x)$. This is a local field, 
proportional to the energy operator $\epsilon(x)$ of the 
thermal Ising model 
\EQ
\Theta(x) = 2 \pi \tau \epsilon(x) \,\,\,, 
\label{deftheta}
\EN 
where $\tau = (T - T_c)$ is the displacement of the temperature 
from its critical value, related to the mass of the model by 
${\cal M} = 2\pi \tau$.  This operator has only a nonzero 
two--particle FF, given by   
\EQ
F^{\Theta}(\beta_1-\beta_2) = 2 \pi {\cal M}^2 \,
\frac{f(\beta_1-\beta_2)}{f(i \pi)} \,\,\,.
\label{FFTheta}
\EN 

The second class consists of the $Z_2$ odd, local or $Z_2$ even, 
semi--local operators, as for instance the magnetization operator 
$\sigma(x)$ and the disorder operator $\mu(x)$. In the 
high--temperature phase of the model, $\sigma(x)$ is an odd, 
local operator. Its FF give rise to the infinite sequence 
$F_{2 n +1}^{\sigma}$ on all odd numbers of external particles. 
Taking into account the kinematical poles at $\beta_{ij} = i \pi$, 
the explicit expression of these FF is given by 
\EQ
F_{2 n +1}^{+} = H_{2 n + 1} \prod_{i<j}^{2 n + 1} 
{\cal F}(\beta_{ij}) 
\,\,\, . 
\label{FFsigma}
\EN 
$H_{2 n +1}$ is a normalization constant, given by $H_{2 n +1} = 
i^n H_1$, where $H_1$ is the one--point FF, 
$
\langle 0 | \sigma(0) |A \rangle = H_1
$.

The disorder field  $\mu(x)$, in the high--temperature 
phase, is an even semi--local field. Its FF extend on an 
infinite sequence $F_{2n}^{\mu}$ on all even numbers 
of asymptotic states and their expression is given by 
\EQ
F_{2 n }^{-} = H_{2 n} \prod_{i<j}^{2 n} {\cal F}(\beta_{ij})
\,\,\, , 
\label{FFmu}
\EN 
with the normalization constant $H_{2 n} =  i^n H_0$ and  
$H_0$ the vacuum expectation value of this operator, 
$ 
\langle 0 | \mu(0) |0 \rangle = H_0 
$.
Switching between the high and low temperature phases, the 
$Z_2$ odd and even fields $\sigma(x)$ and $\mu(x)$ simply 
swap their role. 

The above explicit expressions of the FF allow to obtain
 exact informations 
on the thermal Ising model. We can compute, in 
particular, its correlation functions and related quantities. 
The simplest correlator is given by 
\EQ
G(x) = 
\langle \Theta(x) \Theta(0) \rangle \,\,\, , 
\EN
($ x \equiv {\cal M} R$). By inserting a complete set of 
intermediate states, it can be expressed as 
\begin{eqnarray}
G(x) & = & \frac{1}{2} \int \frac{d\beta_1}{2\pi} 
\frac{d\beta_2}{2\pi} \left|F^{\Theta}(\beta_1 - \beta_2)\right|^2 
\, e^{-x (\cosh\beta_1 + \cosh\beta_2)} = 
\label{Bessel}\\
&= & {\cal M}^4 \left[ K_1^2(x) - K_0^2(x) \right] \,\,\,,
\nonumber 
\end{eqnarray} 
where $K_i(x)$ are the ordinary Bessel functions. In the 
ultraviolet limit $x \rightarrow 0$, the above expression 
reduces to 
\EQ
G(x) \,\simeq\, \frac{{\cal M}^2}{R^2} \,\,\,,
\label{uvising}
\EN 
from which we can read the anomalous dimension of the energy 
operator in its conformal limit, $\eta_{\epsilon} = 1$. 
The knowledge of $G(x)$ also allows to determine the 
central charge of the conformal field theory which arises in 
its ultraviolet limit. To this aim, we can use the $c$--theorem 
sum rule \cite{cth,Cardycth}  
\EQ
C_{uv} = \frac{3}{4} \int d^2x \,|x|^2 \, 
\langle \Theta(x) \Theta(0) \rangle \,\,\, . 
\label{ctheorem}
\EN 
By inserting the explicit expression (\ref{Bessel}) of $G(x)$, 
the above formula gives $C_{uv} = \frac{1}{2}$.

The same set of conformal data can also be extracted by looking 
at the ground--state energy of the theory on an infinite cylinder 
of width $R$. The exact expression is obtained by means of the 
Thermodynamic Bethe Ansatz (TBA) \cite{TBA} and for the 
effective central charge 
\EQ
{\tilde C} \equiv C - 24 \Delta_{\rm min} \,\,\, ,
\label{effective}
\EN  
we have\footnote[1]{In this equation $\gamma_E$ is the
Euler-Mascheroni constant and $\xi(s)$ the Riemann zeta 
function.} \cite{KlMe}
\begin{eqnarray} 
{\tilde C}(x) & = & 
\frac{1}{2} -\frac{3 x^2}{2\pi^2} \left[
\ln\frac{1}{x} + \frac{1}{2} + \ln\pi - \gamma_E \right.
\label{TBAequa}
\\
&& \left. \,\,\,\,\,\, -4 \sum_{n=1}^{\infty} 
\left(\begin{array}{c} \frac{1}{2} \\ n +1\end{array} \right) 
(1 - 2 ^{-2n -1}) \xi(2n + 1) \left(\frac{x^2}{\pi^2}\right)^n   
\right] \,\,\,.\nonumber
\end{eqnarray}
The power series of the above expression should match 
with its evaluation in terms of Conformal Perturbation Theory 
given by  
\EQ
{\tilde C}_{\rm per}(x) = \sum_{n=1}^{\infty} c_n 
\left(\tau R^{2 - \eta_{\epsilon}}\right)^n \,\,\, ,
\label{perturbativecentral}
\EN 
where 
\EQ
c_n = 12 \frac{(-1)^n}{n!} R^{2 (1-n) + n \eta_{\epsilon}} \,
\int_{\rm cyl} \langle \langle \Delta_{\rm min}| 
\epsilon(0) \prod_{j=1}^{n-1} \epsilon(w_j) d^2w_j |
\Delta_{\rm min} \rangle \rangle_{\rm conn} \,\,\,,
\label{pertcoeff}
\EN 
and $\langle\langle ... \rangle \rangle$ denotes 
the conformal correlators on the cylinder. 

Since for the critical Ising model $\Delta_{\rm min} = 0$ and 
the energy operator $\epsilon(r)$ is odd under duality, 
we have $c_n = 0$ for odd $n$. Hence, eq.\,(\ref{TBAequa}) 
fixes the central charge to be $C_{uv} = \frac{1}{2}$. 
Moreover, from the comparison of the power series in (\ref{TBAequa}) 
with the perturbative expansion (\ref{perturbativecentral}), we 
have $\eta_{\epsilon} =1$, in agreement with the determination 
of this quantity obtained by the short--distance limit of the 
correlator (\ref{Bessel}).  

Let us now consider the correlation functions of the $Z_2$ odd, 
local and even, semi--local fields, 
\EQ
\begin{array}{c}
G_-(x) = \langle \sigma(R) \sigma(0) \rangle \,\,\,;\\
G_+(x) = \langle \mu(R) \mu(0) \rangle \,\,\,, 
\end{array}
\EN 
given by 
\EQ
G_{\pm}(x) = \sum_{n}^\infty 
\frac{1}{n!} \int \frac{d\beta_1}{2\pi} \cdots 
\frac{d\beta_n}{2\pi} \left|F^{\pm}_n\right|^2 \, 
e^{-x \sum_{i}^n \cosh\beta_i} \,\,\, ,
\label{FFexpresssigmamu}
\EN 
where the sum runs on odd numbers for $G_+(x)$ and 
on even ones for $G_-(x)$. As shown in \cite{Yurov,McCoy}, 
eq.\,(\ref{FFexpresssigmamu}) provides an integral 
representation of a particular solution of the Painleve' 
equation. An exact analysis of this solution allows, 
in particular, to study its short--distance limit 
$x \rightarrow 0$ with the result 
\EQ
G_{\pm}(x) \,\simeq\, \frac{{\rm const}}{R^{\frac{1}{4}}} \,\,\, . 
\label{uvsigmamu}
\EN 
The above expression determines the anomalous
dimensions of the $Z_2$ odd fields to be $\eta_{\sigma} 
= \eta_{\mu} = \frac{1}{8}$. 

The anomalous dimension of these fields can be equivalently 
extracted by means of the $\eta$ sum--rule \cite{cluster1}
\EQ
\eta_{\Phi} = - \frac{1}{2 \pi \langle \Phi\rangle} \,
\int d^2x \,\langle \Theta(x) \Phi(0) \rangle \,\,\,.
\label{Deltasumrule}
\EN 
This formula can be applied in the high--temperature phase to 
obtain the anomalous dimension of the disorder field and in 
the low--temperature to obtain the anomalous dimension of 
the magnetization operator. By inserting a complete set of 
states between the two operators and noticing that the FF of 
$\Theta(x)$ is different from zero only on the two--particle 
state, we have 
\EQ
\eta_{\mu} = \eta_{\sigma} = \frac{1}{2\pi} 
\int_0^{\infty} \frac{d\beta}{\cosh^2\beta} f(2\beta) {\cal F}(2\beta) = 
\frac{1}{8} \,\,\,.
\label{delsumrule18}
\EN 

It is worth mentioning that the anomalous dimension of these fields 
can be also obtained by exploiting the formal analogy of expression
(\ref{FFexpresssigmamu}) with a grand--canonical partition function
$\Xi(z,L)$ of a fictitious one--dimensional gas in a box of length $L
\sim \log\frac{2}{{\cal M} R}$ and fugacity $z = \frac{1}
{2 \pi}$ \cite{Yurov,CMform}. Hence, by the state equation
of this gas
\EQ
\Xi(z,L) = e^{p(z) L} \,\simeq\, \left(\frac{1}{{\cal M} R}\right)^{p(z)}
\,\,\,, \label{equationstate}
\EN
the anomalous dimension is nothing but its pressure $p(z)$
at $z=\frac{1}{2\pi}$. In the nearest neighborhood approximation
of $\Xi(z,L)$, the pressure $p$ can be obtained as a solution
of the integral equation
\EQ
2 \pi = \int_0^{+\infty} dz e^{-p\,z} \left|{\cal F}_2(z)\right|^2
\,\,\, .
\label{onedimensionalgas}
\EN
In the thermal Ising model, the above formula provides an excellent
approximation of $\eta$, i.e. $\eta_{\sigma} \sim 0.12529$
\cite{CMform}.

\subsection{Form Factors for the Elliptic Ising Model}

Let us discuss now the Form Factors of the theory described by the 
$S$--matrix $S = -1$  but this time regarded as a periodic function 
of $\beta$. As it is clear from eq.\,(\ref{usefulidentity}), 
there is a complete arbitrariness in the choice of the periodic 
realization of $-1$. Instead of working with the $S$--matrix 
(\ref{Ising}) of period $T$, it is more convenient to consider 
the following $2 T$ periodic realization 
\EQ 
-1 = \exp\left[-4 i
\,\sum_{n=1}^{+\infty} \frac{1}{2 n -1} 
\sin\left(\frac{\pi (2 n - 1) \beta}{T}\right) 
\right] \,\,\, , 
\label{Ising2T} 
\EN 
which leads to a simplification in the expressions of the Form 
Factors of the model, without altering its physical content
\footnote[2]{
In fact, one can always convert the expressions of the FF relative 
to the period $2T$ to those relative to period $T$ by multiplying 
them for even, $2\pi i$ periodic functions which enter the general 
solution of (\ref{cond}). For instance, the function 
$\sn{\left(i\frac{\ka'}{T}\beta\right)}$, solution of (\ref{cond}) 
with period $2 T$, can be transformed by means of the even, 
$2\pi i$ periodic function 
$\left[l \,\sn^2{\left(i\frac{\ka'}{T}\beta\right)} + 1\right]$
into the $T$--periodic solution 
$\frac{\sn{\left(i\frac{\ka'}{T}\beta\right)}}
{l \,\sn^2{\left(i\frac{\ka'}{T}\beta\right)} + 1}$ 
of the same equation.}. 

In what follows we will not address the problem of finding the 
most general solutions of all FF equations relative to this 
theory. We will rather concentrate our attention on particular 
classes of solutions which have the minimal analytic structure 
compatible with all the constraints. 

Let us consider first the minimal two--particle Form Factors,
solutions of the equations 
\bea
& F(\beta) ~=~ - \, F(-\beta) \,\,\,;\nonumber \\
& F(i\pi - \beta) ~ =  e^{2\pi i\omega} ~ F(i\pi + \beta) \,\,\, . 
\label{cond}
\ena
They are determined up to arbitrary even, $2\pi i$--periodic and 
$2T$ (anti)--periodic, analytic functions of $\beta$. This time we are 
looking for the simplest solution of these equations in the space of 
elliptic functions. Using the properties of the Jacobian elliptic 
functions listed in Appendix A, it is easy to see that they are 
given by  
\EQ
F_{\rm min}(\beta) = \left\{
\begin{array}{cc}
f(\beta) = -i \,
\frac{\sn\left(i \frac{\ka'}{T} \beta\right)}
{\dn\left(i \frac{\ka'}{T} \beta\right)} \,\,\, ,
 & \,\,\, \mbox{if $\omega = 0$} \,\,\, ;\\
  & \\
{\cal F}(\beta) = - i\,
\frac{\sn\left(i \frac{\ka'}{T} \beta\right)}
{\cn\left(i \frac{\ka'}{T} \beta\right)} \,\,\, ,& 
\,\,\, \mbox{if $\omega = \frac{1}{2}$} 
\,\,\,.
\end{array}
\right.
\label{FminEllIsing}
\EN 
These expressions represent the simplest generalization of the 
standard results (\ref{FminIsing}) with no poles on the real axis 
and which accounts for the real periodicity of the system under 
consideration, with the property $F_{\rm min}(\beta + 2 T) = - 
F_{\rm min}(\beta)$. Their analytic structure consists of 
the following set of simple zeros and poles: 

\EQ
\begin{array}{ll} 
f(\beta) \longrightarrow & \left\{
\begin{array}{ll}
&{\rm Zeros:}~~~~
\beta_{m,n} = ~ 2 m \pi i + 2 n T \,\,\,;\\
&{\rm Poles:}~~~~
\beta_{m,n} = ~ (2 m +1) \pi i + (2 n+1) T \,\,\, ;
\end{array}
\right. \\
  & \\
{\cal F}(\beta) \longrightarrow & \left\{
\begin{array}{ll}
&{\rm Zeros:}~~~~
\beta_{m,n} = ~ 2 m \pi i + 2 n T \,\,\,;\\
&{\rm Poles:}~~~~
\beta_{m,n} = ~ (2 m +1) \pi i + 2 n T \,\,\, , 
\end{array}
\right.
\end{array}
\label{analyticFFFF2}
\EN
with $m,n \in {\bf Z}$.
Notice that $f(\beta)$ has an infinite sequence of poles 
at $\beta = i \pi + (2 n +1) T$, located just at the edge of 
the second sheet. They correspond to the infinite sequence 
of the one--particle intermediate states given by the resonances 
$R_n$ (see the discussion at the end of Section 2). On the 
contrary, the poles of ${\cal F}(\beta)$ located at 
$\beta = i\pi +2nT$ do not signal the presence of resonances but
are simply related to the infinitely repeated pattern of the branch 
cut of the $t$--channel of the $S$--matrix. 

Concerning the recursive equations satisfied by the FF, they 
are given by the same equation (\ref{recursiveIsing}) and 
therefore, as before, the operators fall into two different 
sectors. 

Let us consider first the Form Factors of the local $Z_2$ 
even or semi--local odd operators, i.e. the ones which do 
not have poles at $\beta_{ij} = i \pi$. Their general 
expression can be written in terms of the function $f(\beta)$ 
and they are nonzero only on a finite number of external 
states. Let us consider, in particular, the FF of the trace 
operator. Even in the periodic case we assume this operator to be 
related to the energy field by eq.\,(\ref{deftheta}). Its 
only non--vanishing FF is the one on the two--particle state  
which we assume to be 
\EQ
F^{\Theta}(\beta_1-\beta_2) 
\equiv \langle 0 | \Theta(0) | \beta_1 \beta_2 \rangle
~=~ 2 \pi {\cal M}^2 \,\frac{f(\beta_1 - \beta_2)}{f(i \pi)}\,\,\,.
\label{FFF}
\EN 
For the two--point function of this operator we have 
\begin{eqnarray}
G(x) & = & \langle \Theta(R) \Theta(0) \rangle = \\
& & \, = \frac{1}{2} \int \frac{d\beta_1}{2 \pi} 
\frac{d\beta_2}{2 \pi} \left|F^{\Theta}(\beta_1 - \beta_2)\right|^2 
e^{-x (\cosh\beta_1 + \cosh\beta_2)} \,\,\,.\nonumber 
\end{eqnarray}
By using the variables $y_{\pm} =\frac{1}{2}(\beta_1 \pm \beta_2)$, 
it can be written as 
\EQ
G(x) = \frac{4 {\cal M}^4}{|f(i\pi)|^2}
\,\int_0^{+\infty} dy_- |f(2 y_-)|^2 K_0(2 x \cosh y_-) \,\,\,.
\label{finaleelliptic}
\EN
Plots of this correlation function for different values of 
the modulus $l$ are shown in Figure 7. 

Let us discuss in more detail the large and the short distances 
behaviour of this correlator. As evident from Figure 7, the large 
distance behaviour of the correlator (\ref{finaleelliptic}) is 
essentially the same for all value of the period $T$ and is ruled 
by the mass gap ${\cal M}$ of the stable particle. This is 
confirmed by the analytic estimate of $G(x)$ for large values of 
$x$: the only parameter which contains informations on the periodic 
structure of the theory (in terms of parameters of the elliptic 
functions) is the normalization in front of the exponential decay 
\EQ
G(x) \simeq 
\frac{2}{\pi} \left({\cal M}^2\,l'\,\ka\right)^2 \,
\frac{e^{-2 x}}{x^2} + {\cal O}\left(e^{-4 x}\right) \,\,\,.
\label{asymptotic}
\EN 
The situation is drastically different in 
the ultraviolet regime. 
While in the non--periodic case ($l=0$) the two--particle FF is 
expressed by the unbounded function $\sinh\frac{\beta}{2}$, 
for $l \neq 0$ it is instead a periodic, regular and 
limited function on the real axis which can always be bounded 
by a constant $B$, leading to\footnote[3]{
To obtain this result we have used eq.(6.663.1) of \cite{Gran}.}
\EQ 
G(x) \leq 4 {\cal M}^4 B^2 \int_0^{+\infty} dy K_0(2 x \cosh y) = 
2 \left({\cal M}^2 B \right)^2 \, K_0^2(x) \,\,\, . 
\label{exactint}
\EN 
This inequality implies that for $ l \neq 0$ $G(x)$ cannot have a 
power law singularity in its ultraviolet limit as, instead, 
it happens in the
standard Ising Model (see eq. (\ref{uvising})). Hence the conclusion 
is simply that the anomalous dimension of the energy operator changes
discontinously from the value $\eta_{\epsilon} =1$ (for $l=0$) 
to $\eta_{\epsilon} = 0$ for all other finite values of the module! 

This result seems to give rise to an apparent paradox. In 
fact, one could argue that the anomalous dimension of the 
energy operator can be equivalently extracted in terms of 
the TBA equations by looking at the perturbative part of 
the ground state energy. Since this quantity is still given 
by the same eq.\,(\ref{TBAequa}), it appears that from the TBA 
approach one would rather reach the previous conclusion, i.e. 
$\eta_{\epsilon} = 1$. How can we conciliate this apparent 
mismatch of the values of $\eta_{\epsilon}$ obtained by the 
two different methods? 

The solution to this puzzle is the following. First of all, notice 
that the TBA does not compute the central charge of the theory 
but rather provides the evaluation of the effective central 
charge of the theory, ${\tilde C} = C - 24 \Delta_{\rm min}$. To
compute directly the central charge, we have instead to employ 
the $c$--theorem sum--rule (\ref{ctheorem}). The result for 
the central charge of the Elliptic Ising  Model is reported in 
Figure 8. This figure shows that, by varying the period, the 
central charge of the model is no longer $C = \frac{1}{2}$ but, 
on the contrary, takes any value between $0$ and $\frac{1}{2}$, 
depending on the module $l$. On the other hand, since the effective 
central charge ${\tilde C}$ is always equal to $\frac{1}{2}$ for any 
value of the module $l$, there should be in the periodic 
realization of the Ising model an operator with conformal 
weight $\Delta_{\rm min} < 0$. Hence, the underlying conformal 
theory of the Elliptic Ising Model is no longer unitary (and rational) 
and consequently we have no reasons to argue that the 
perturbative coefficients $c_n$ vanish for $n$ odd, as in the Ising 
Model. Therefore, 
the matching of the Conformal Perturbation Theory in the 
presence of a field with $\Delta_{\rm min} \neq 0$ with the 
corresponding perturbative part of (\ref{TBAequa}) gives 
$\eta_{\epsilon} = 0$, in agreement with the determination 
done by looking at the ultraviolet limit of the correlator. 
For the limiting value $l =1$, we have $\Delta_{\rm min} = 
- \frac{1}{48}$. 

As already discussed, the FF of even, local operators inherit 
from the $S$--matrix
the pole structure associated to the resonance states. 
These states are even under the $Z_2$ symmetry of the model 
and therefore they couple directly to the operator $\Theta$. We can 
extract the matrix elements $\langle 0 |\Theta(0)| R_n\rangle$ by 
formally applying the bound state residue equation (\ref{respole}), 
with $\Gamma_{AA}^{R_n} = i {\sqrt {\pi\delta}}$ ($\delta \rightarrow 0$). 
As a result we have 
\EQ
(-1)^n \,\frac{2\pi^2 {\cal M}^2}{l\,\ka} = 
{\sqrt{\pi \delta}} \, \langle 0 |\Theta(0) | R_n \rangle \,\,\,.  
\label{onepointfunction}
\EN  
Hence, $\langle 0 | \Theta(0) | R_n\rangle$ diverges as 
$1/{\sqrt \delta}$ but its product with the three-particle vertex 
of the resonances is constant in the limit $\delta \rightarrow 0$. 

Let us consider now the FF of the $Z_2$ odd, local and even, 
semi--local fields, concentrating our attention on the 
magnetization and disorder operators. The magnetization 
$\sigma(x)$ is local and has FF on all odd 
numbers of external particles. A close solution of the 
recursive equations (\ref{recursiveIsing}) 
is given by 
\EQ
F^{+}_{2n + 1} = \left(\frac{2 \ka i}{\pi}\right)^n 
\, (l')^{n (n+1)} H_1 \,\prod_{i<j}^{2 n+1} 
{\cal F}(\beta_{ij}) \,\,\,, 
\label{FFElsigma}
\EN 
where $H_1 = \langle 0 |\sigma(0)|A \rangle$. 

The disorder operator $\mu(x)$ is semi--local and has FF 
on all even numbers of external particles 
\EQ
F^{-}_{2n} = \left(\frac{2 \ka i}{\pi}\right)^n 
\, (l')^{n^2} H_0 \,\prod_{i<j}^{2 n} {\cal F}(\beta_{ij}) \,\,\,, 
\label{FFElmu}
\EN 
where $H_0 = \langle 0 | \mu(0) | 0 \rangle$. All the FF 
(\ref{FFElsigma}) and (\ref{FFElmu}) possess the kinematic pole at 
$\beta_{ij} = i \pi$ as well as at their periodic repeated 
positions $\beta_{ij}^{(n)} = i \pi + 2 n T$. 

For the two--point correlation functions of these fields we have 
\EQ
\begin{array}{c}
{\cal G}_-(x) = \langle \sigma(R) \sigma(0) \rangle \,\,\,;\\
{\cal G}_+(x) = \langle \mu(R) \mu(0) \rangle \,\,\,, 
\end{array}
\label{correlatorElsigma}
\EN 
with 
\EQ
{\cal G}_{\pm}(x) = \sum_{n}^\infty 
\frac{1}{n!} \int \frac{d\beta_1}{2\pi} \cdots 
\frac{d\beta_n}{2\pi} \left|F^{\pm}_n\right|^2 \, 
e^{-x \sum_{i}^n \cosh\beta_i} \,\,\, ,
\label{FFElexpresssigmamu}
\EN 
where the sum runs on odd numbers for ${\cal G}_+(x)$ and 
on even ones for ${\cal G}_-(x)$. 

As for the energy field, at large distances  
these correlation functions have an exponential decay, 
ruled by the mass gap ${\cal M}$ of the stable part of the 
spectrum 
\EQ
{\cal G}_-(x) \,\simeq\, H_1^2 \sqrt{\frac{1}{2\pi x}}\,e^{-x} 
+ {\cal O}\left(e^{-3 x}\right) \,\,\,,
\label{expsigma}
\EN 
\[
\hspace{10mm}
{\cal G}_+(x) \,\simeq\, 
\frac{2}{\pi^5} \left({\it l}'\, \,\ka^2\,H_0\right)^2 \,
\frac{e^{-2 x}}{x^2} + {\cal O}\left(e^{-4 x}\right) \,\,\,.
\]
The only informations on the periodic structure of the theory 
may enter the normalization in front of the exponentials. 
Let us now analyse the short--distance behaviour of 
${\cal G}_{\pm}(x)$. In absence of an exact resummation of 
the series (\ref{FFElexpresssigmamu}), 
we have to rely on other methods for 
computing the anomalous dimensions of the corresponding fields. 
It is easy to see that, in contradistinction to the anomalous 
dimension $\eta_{\epsilon}$ of the energy operator which jumps 
discontinuosly from 
$1$ to $0$ by introducing a finite period $T$, the anomalous 
dimension $\eta_{\sigma}$ (or equivalently $\eta_{\mu}$) depends
continuously on $T$. In fact, 
in the ultraviolet limit we can apply the thermodynamics 
approach as in eq. (\ref{equationstate}) 
and write the following representation 
for the correlator 
\EQ
{\cal G} = {\cal G}_+ + {\cal G}_- \,\simeq \, 
\left(\frac{1}{{\cal M} R}\right)^{p(l)} \,\,\,,
\label{equationstate2}
\EN 
where the pressure of the one--dimensional fictitious gas is given
in eq. (\ref{onedimensionalgas}). Since in this case  
$|{\cal F}(\beta)|^2$ is a regular and bounded function along 
the real axis for {\em any} value of the module $l$, it follows  
that $p(l)$ is a continous function of the module. 
Another way to reach the same conclusions is to analyse the exact expression 
for the anomalous dimensions as obtained by the $\eta$ 
sum--rule
\EQ
\eta_{\sigma}(l) = \eta_{\mu}(l) = 
\frac{1}{2\pi} \int_0^{\infty} \frac{d\beta}{\cosh^2\beta}
\frac{F_2^-(2\beta)}{H_0}  \,\frac{f(2\beta)}{f(i\pi)} \,\,\,.
\EN
The relative graph of this function is reported in Figure 9. 
The curve starts from $\eta=\frac{1}{8}$ at $l=0$ and goes 
then continously to zero at $l=1$.  

\resection{Form Factors for the Elliptic Sinh--Gordon Model}

Let us now consider the quantum field theory described by the 
elliptic $S$--matrix (\ref{ellipS}) and address in its full 
generality the problem of computing exactly its FF. We will see 
that the FF equations admit in this case a large class of 
solutions which for $l \rightarrow 0$ all collapse to the 
same solutions of the Sinh--Gordon model. As a matter of fact, 
due to the absence of constraints on the asymptotic behaviour of FF, 
the final set of solutions has indeed infinite dimension.
 
We concentrate only on the case of FF of local operators. 
As usual, the first step consists in the evaluation of the 
two--particle Form Factor. For any local operator,
this is a solution of the monodromy eqs.\,(\ref{F2}) with 
$\omega = 0$. Using the exponential representation (\ref{secondexp}) 
for the $S$--matrix and the formula (\ref{fourierFFexp}) we have 
\EQ
F_{\rm min}(\beta) ~=~ {\cal N}\,\exp \left[4 \sum_{n=1}^{\infty} 
\frac{1}{n} \frac{\sinh\left(\frac{n{\it a} \pi^2}{T}\right) \sinh\left(
\frac{n(1-{\it a}) \pi^2}{T}\right)}{\cosh\left(\frac{n \pi^2}{T}\right)  
\sinh\left(\frac{2n \pi^2}{T}\right)} \sin^2 \left(n \frac{\pi}{T} 
\hat{\beta} \right)\right] \,\,\, ,
\label{ESGmin}
\EN
where $\hat{\beta} \equiv i\pi - \beta$ and ${\cal N}$ is a 
normalization constant. This solution satisfies the periodicity 
condition $F_{\rm min}(\beta+T) ~=~ F_{\rm min}(\beta)$. 
The previous expression for $F_{\rm min}$ can be further elaborated 
in order to get a final form, more suitable for the determination of 
its analytic structure and for explicit calculations. As described in 
details in Appendix C, it can be rewritten, up to a constant, as an 
infinite product of trigonometric functions  
\EQ
F_{\rm min}(\beta) ~=~
\prod_{k=2}^{\infty} \left( W(\beta,k) \right)^{k-1} \,\,\, ,
\label{ESGmin2}
\EN 
where 
\[
W(\beta,k) = \frac{
\left(1-2q_6 \cos\left(2\frac{\pi}{T} \hat{\beta}\right) +q_6^2\right)
\left(1-2q_5 \cos\left(2\frac{\pi}{T} \hat{\beta}\right) +q_5^2\right)
\left(1-2q_4 \cos\left(2\frac{\pi}{T} \hat{\beta}\right) +q_4^2\right)}
{\left(1-2q_1 \cos\left(2\frac{\pi}{T} \hat{\beta}\right) +q_1^2\right)
\left(1-2q_2 \cos\left(2\frac{\pi}{T} \hat{\beta}\right) +q_2^2\right)
\left(1-2q_3 \cos\left(2\frac{\pi}{T} \hat{\beta}\right) +q_3^2\right)}
\]
and $q_i=q_i(k)$, $i=1, \cdots,6$ are the factors listed in 
eq.\,(\ref{qfactors}). Using the identity (\ref{usefulidentity13}) 
we can also express the previous result as an infinite product 
of gamma functions 
\EQ
F_{\rm min}(\beta) = \prod_{n=-\infty}^{+\infty} \,
\prod_{k=0}^{+\infty} Y_{k,n}(\beta) \,\,\,,
\label{anotherexpression}
\EN 
where 
\[
Y_{k,n}(\beta) = 
\left|
\frac{\Gamma\left(k+\frac{3}{2}+\frac{i\hat\beta_n}{2\pi}\right)
\Gamma\left(k+\frac{1}{2}+\frac{a}{2}+\frac{i\hat\beta_n}{2\pi}\right)
\Gamma\left(k+1-\frac{a}{2}+\frac{i\hat\beta_n}{2\pi}\right)}
{\Gamma\left(k+\frac{1}{2}+\frac{i\hat\beta_n}{2\pi}\right)
\Gamma\left(k+\frac{3}{2}-\frac{a}{2}+\frac{i\hat\beta_n}{2\pi}\right)
\Gamma\left(k+1+\frac{a}{2}+\frac{i\hat\beta_n}{2\pi}\right)}
\right|^2 \,\,\,,
\] 
and $\hat\beta_n = \hat\beta + n T$. 

The solution (\ref{ESGmin2}) is analytic in the $\beta$--plane  
except for an infinite set of poles (see Appendix C for details)
\EQ
\begin{array}{llllll}
{\rm Poles:} ~~~ &\beta_n(k) ~=~ -i\pi a~+~ (2k-1)i\pi  ~+~ nT \,\,\,;
\nonumber \\
&\beta_n(k) ~=~  -i\pi a~+~ (2-k) 2i\pi  ~+~ nT \,\,\,;
\nonumber \\
&\beta_n(k) ~=~ i\pi a~+~ (3-2k)i\pi  ~+~ nT \,\,\,;
\nonumber \\
&\beta_n(k) ~=~ i\pi a ~+~  (k-1)2i\pi ~+~ nT 
\,\,\,;
\nonumber \\
&\beta_n(k) ~=~ 2ki\pi  ~+~ nT \,\,\,;
\nonumber \\
&\beta_n(k) ~=~ (1-k)2i\pi  ~+~ nT \,\,\, ,
\label{ESGpoles}
\end{array}
\EN
and it vanishes at the following locations
\EQ
\begin{array}{llllll}
{\rm Zeros:} ~~~ &\beta_n(k) ~=~ -i\pi a~+~ 2ki\pi ~+~ nT 
\,\,\,;
\nonumber \\
&\beta_n(k) ~=~ -i\pi a ~+~(3-2k)i\pi ~+~ nT 
\,\,\,;
\nonumber \\
&\beta_n(k) ~=~ i\pi a ~+~ (1-k)2i\pi ~+~ nT 
\,\,\,;
\nonumber \\
&\beta_n(k) ~=~ i\pi a ~+~(2k-1)i\pi ~+~ nT 
\,\,\,;
\nonumber \\
&\beta_n(k) ~=~ (k-1)2i\pi ~+~ nT 
\,\,\,;
\nonumber \\
&\beta_n(k) ~=~ (2-k)2i\pi ~+~ nT \,\,\,.
\label{ESGzeros}
\end{array}
\EN
For given $k$, poles and zeros have multiplicity $(k-1)$ and they are 
drawn in Figure 10. The minimal solution $F_{\rm min}$ does not have
singularities in the physical sheet but only single zeros in $\beta=nT$.
The poles at $\beta = -i \pi a + n T$ on the unphysical sheet 
of $F_{\rm min}(\beta)$ may be regarded as ``bound state'' poles 
of the FF due to the resonances. 

The solution (\ref{ESGmin2}) satisfies the functional relation
\EQ
F_{\rm min}(\beta+i\pi) \, F_{\rm min}(\beta) ~=~ 
\frac{\sn\left(2 i \frac{\ka'}{T}\beta \right)}{\sn\left(
2i \frac{\ka'}{T}\beta \right)~-~ \sn\left(2\ka {\it a}\right)} \,\,\,.
\label{relation}
\EN
This can be easily proven by studying the distribution of poles
and zeros for the product on the l.h.s., as shown in Figure 11. 
When $F_{\rm min}(\beta)$ is multiplied by the same function 
evaluated at the shifted point $(\beta+i\pi)$ a partial
cancellation of poles against zeros occurs, leaving only a 
distribution of simple poles and zeros (third line of Figure 11). 
This distribution indeed corresponds to the pole singularities 
and zeros of the elliptic function on the r.h.s. of 
eq.\,(\ref{relation}). Notice that we have used the functional 
equation (\ref{relation}) as it stands in order to fix a 
convenient normalization ${\cal N}$ of $F_{\rm min}(\beta)$. 

In order to simplify the following equations, it is convenient 
to introduce the notation 
\EQ
{\rm u}\equiv i\frac{\ka'}{T} \beta \,\,\,\,\,\,,\,\,\,\,
{\rm u}_{ij}\equiv i\frac{\ka'}{T} \beta_{ij} \,\,\,. 
\label{notation}
\EN 

The general structure of $n$--particle Form Factors is given in 
eq.\,(\ref{parametrization}) in terms of the minimal solution 
(\ref{ESGmin2}). Since the theory has no bound states, the only 
poles present are the kinematic poles associated to every 
three--particle subchannels ($\beta_{ij} =i\pi$). This suggests 
to parameterise the functions $K_n$ entering the general expression 
of the FF (\ref{parametrization}) as follows
\EQ
K_n(\beta_1,\cdots, \beta_n) ~=~ H_n \, 
Q_n(\beta_1,\cdots, \beta_n)\,
\prod_{i < j} \frac{1}
{\cn \,u_{{\it ij}}}
\,\,\, ,
\label{FFn}
\EN
where $H_n$ are normalization constants. Notice that in the
above expression the presence of the kinematic pole has been 
made explicit by the introduction of the periodic elliptic 
functions $\cn \,u_{{\it ij}}$. The choice of this parameterization 
is suggested by the general requirement that the residue of $F_n$ 
at the kinematic pole has to be a periodic function of the other 
$(n-2)$ rapidity variables (see the general discussion in Sect. 3). 
Therefore, the kinematic pole at $\beta_{ij} =i\pi$ appears toghether 
with an infinite sequence of poles located at $\beta_{ij} = i\pi + 2nT$. 
Even though the function $\cn \,u$ is $2 T$ periodic, the actual 
$T$--periodicity of the residue of $F_n(\beta_1,...\beta_n)$ 
at $\beta_{ij} = i\pi$ will be restored by the consequent expressions 
of $Q_n(\beta_1,...\beta_n)$ obtained by using this parameterization.

The function $\cn [u(\beta)]$ satisfies the identity 
\EQ
\cn [u(\beta + i \pi)] 
\,\cn [u(\beta)] = 
{\it l}' \,\frac{ \sn \,u\, \cn \,u}{
\dn \,u } \,\,\,.
\label{Grelation}
\EN
The terms $Q_n$ are symmetric, $2\pi i$ periodic
functions of the $\beta_i$'s, which have to be determined by 
the recursive equations (\ref{recursive}) satisfied by 
the FF. By using the expression (\ref{2ellipS}) for the $S$--matrix 
and exploiting the functional relations (\ref{relation}) and 
(\ref{Grelation}) together with eq.\,(\ref{1+dn2}), we obtain the 
following recursive equations for the functions $Q_n$
\EQ
Q_{n+2} (\beta+i\pi, \beta, \beta_1, \cdots ,\beta_n) ~=~
{\cal D}_n(\beta \,| \,\beta_1, \cdots ,\beta_n) \,
Q_n (\beta_1, \cdots ,\beta_n)   
\label{recursive2}
\EN
where 
\EQ
{\cal D}_n(\beta \,| \,\beta_1, \cdots ,\beta_n) ~=~ 
\frac{(- 2 i)^n}{2 i \,\sn\left(2 \ka {\it a}\right)} \times 
\label{functionD}
\EN
\[
\frac{
\prod_{i=1}^{n} 
\left[ \sn[2(u -u_{\it i})]
- \sn \left(2\ka {\it a}\right) \right] 
- \prod_{i=1}^{n} 
\left[ \sn[2(u -u_{\it i})] + \sn \left(2\ka {\it a}\right) 
\right]}
{\prod_{i=1}^n \left[1 + \dn[2 (u - u_{\it i})]\right]} \,\,\,.
\]
Concerning the normalization constants $H_n$, they have been
conveniently chosen as
\EQ
H_{2n} ~=~
\left(\frac{i l'}{2}\right)^{n(n-1)} \, 
\left(\frac{-2 i l' \,\ka \,\sn\left(2 \ka {\it a}\right)}
{\pi \, F_{\rm min}(i\pi)}\right)^{n} \, H_0 \,\,\, ,
\EN
\EQ
H_{2n+1} ~=~ 
\left(\frac{i l'}{2}\right)^{n^2} \, 
\left(\frac{-2 i l' \,\ka \,\sn\left(2 \ka {\it a}\right)}
{\pi \, F_{\rm min}(i\pi)}\right)^{n} \, H_1 \,\,\, ,
\EN 
with the constants $H_0$, $H_1$ determined by the vacuum 
and the one--particle FF of the operator ${\cal O}$. 

Let us discuss the general properties of the recursive equations 
(\ref{recursive2}). It is easy to see that the solutions $Q_n$ 
with $n > 1$ span in this case an infinite dimensional vector space. 
In fact, if $\tilde{Q}_n(\beta_1, ..., \beta_n)$ is a non--trivial 
solution of the recursive eq. (\ref{recursive2}), the general 
expression 
\EQ
Q_n(\beta_1, ..., \beta_n) ~=~ 
\tilde{Q}_n(\beta_1, ..., \beta_n)
~+~ A_n(\beta_1, ...,\beta_n) \, W(\beta_1,...,\beta_n)
\label{generalsol}
\EN
is still a solution if $W(\beta_1,...,\beta_n)$ belongs 
to the kernel of the recursive equation 
\EQ
W_n(\beta + i\pi,\beta,\beta_3,...,\beta_n) = 0 \,\,\, , 
\label{kernel}
\EN 
i.e. $W_n(\beta_1,...,\beta_n) = \prod_{i<j} \cn \,u_{{\it ij}}$, and
$A_n(\beta_1, \cdots,\beta_n)$ is a generic $2\pi i$ 
periodic, symmetric function with possible pole singularities
outside the physical strip and which reduces to a constant in the 
limit $l \rightarrow 0$ -- the last requirement in order to reproduce 
the standard result for the Sinh--Gordon model\footnote[4]{In the
Sinh--Gordon case this a--priori infinite set of solutions is severely
restricted by the constraints on the asymptotic behaviour of $F_n$ which
eventually force the function $A_n$ to be a constant \cite{KM}.}. 
Due to the periodic nature of the theory under consideration, no extra
constraints can be imposed on the function $A_n$ which then spans an
infinite space of solutions. Moreover, as we will see from explicit 
examples below, there exist in general several different non--trivial 
solutions $\tilde{Q}_n$ which however reduce to the same solution 
of the FF equations of the Sinh--Gordon model in the limit $l 
\rightarrow 0$. In other terms, given a solution $\tilde{Q}_n$ 
of the FF equations of the Sinh--Gordon model, there are in 
general many different ways to extend this function for $l \neq 0$ 
and obtain then distinct solutions of the FF equation of the 
Elliptic Sinh--Gordon model. 
 
Let us concentrate now on the construction of particular solutions $Q_n$
of the recursive equations.

For $n=1$, $Q_1$ is a constant, as required by relativistic invariance. 
By absorbing this constant into the normalization $H_1$, we can take 
$Q_1 =1$. 

For $n=2$ a solution is given by 
\EQ
Q_2(\beta_1,\beta_2) = 
\cn \,u_{12}  \,\,\,. 
\label{Q2}
\EN
In fact, the two--particle FF of any local operator cannot have a 
pole at $\beta_{12} = i \pi$ so that $Q_2(\beta_1,\beta_2)$ must 
precisely cancel the pole introduced by the function $\cn u_{12}$ 
in the parameterization (\ref{FFn}). For $l \rightarrow 0$ this 
solution reduces to the corresponding one for the Sinh--Gordon model
\cite{FMS,KM}. However, this is not the only acceptable solution 
with this property. We can for instance divide the function (\ref{Q2}) 
by $\dn \,u_{12}$ and obtain another admissible two--particle FF, 
with no pole at $\beta_{12} = i \pi$ and which reduces to the same 
function of the Sinh--Gordon model in the limit $l \rightarrow 0$. 
This new solution differs from the previous one by the distribution of
zeros and poles on the complex $\beta$--plane. In particular,  
it possesses poles at $\beta_{12} = (2 m +1) i \pi + 
(2 n + 1) T$ and zeros at $\beta_{12} = 2 i m  \pi + (2 n + 1) T$, 
which are pushed to infinity for $l \rightarrow 0$. 

Along these lines, it is easy to see that the most general 
solution to the recursive equations for $n=2$ falls into 
the class (\ref{generalsol}), i.e. it may be written as 
\EQ
Q_2(\beta_1,\beta_2) = 
\cn \,u _{12} \,\, q(\beta_{12},l) \,\,\,,  
\label{Q222}
\EN
where $q(\beta_{12},l)$ is any $2\pi i$ periodic, even function 
with no pole at $\beta_{12} = i \pi$ and such that $\lim_{l\rightarrow 0} 
q(\beta_{12},l) = {\rm constant}$. Different solutions will have 
different distributions of zeros and poles but, in the absence 
of extra requirements on the FF of this theory, they are all acceptable, 
as far as they do not possess dangerous poles on the physical strip.  

Let us now discuss the three--particle FF. The function  
$Q_3(\beta_1,\beta_2,\beta_3)$ must be a solution of the
equations (\ref{recursive2}) specialized to the case $n=3$
\EQ
Q_3 (\beta+i\pi,\beta,\beta_3) ~=~ \frac{2}{1 + 
\dn[2 (u - u_3)]} \,\,\, ,
\label{Q3recursive}
\EN
where we have used $Q_1 =1$, absorbing the constant in $H_1$. 
It is easy to find two different non--trivial solutions of this 
equation with no poles in the physical sheet of each two--particle 
subcluster. The first one is given by 
\begin{eqnarray}
Q_3^{(1)}(\beta_1,\beta_2,\beta_3) & =& 
-\left[ 
\frac{\cn (u_{12}+u_{13}) \, \cn \,u_{23}}
{\dn (u_{12}+u_{13}) \, \dn \,u_{23}} +
\frac{\cn (u_{23} + u_{21}) \,\cn \,u_{13}}
{\dn (u_{23}+u_{21}) \, \dn \,u_{13}} ~+~ 
\right. \nonumber 
\\
&& \,\,\,\,\,\,\,\left. + ~
\frac{\cn (u_{13} + u_{23}) \,\cn \,u_{12}}
{\dn (u_{13}+u_{23}) \,\dn \,u_{12}} \right]
\,\,\,. 
\label{Q3solution1} 
\end{eqnarray}
The second solution is given by 
\EQ
Q_3^{(2)}(\beta_1, \beta_2, \beta_3) = \left[
\frac{\cn^2 \,u_{12}}{
\dn^2\,u_{12}} +
\frac{\cn^2\,u_{13}}{
\dn^2\,u_{13}} +
\frac{\cn^2\,u_{23}}{
\dn^2\,u_{23}}\right] \,\,\,.
\label{Q3solution2}
\EN
Both expressions for $T \rightarrow \infty$ fall into the general 
$Q_3$ solution for the Sinh--Gordon model which is given by  
\cite{KM}.
\EQ
Q_3^{Sh}(\beta_1,\beta_2,\beta_3) = A_1(\cosh\beta_{12} 
+ \cosh\beta_{13} + \cosh\beta_{23}) + A_2 \,\,\,,
\label{3Sh}
\EN 
with $A_2 - A_1 = 1$. We notice that both the solutions are periodic in
each variable, with period $2T$, even if they satisfy the condition for the
residue at the kinematic pole to be $T$--periodic, as it follows from the
recursive equation (\ref{recursive}). However, in general nothing prevents
one from finding also $T$--periodic solutions. 

As in the $n=2$ case we can generate a {\em pletora} of solutions 
by multiplying the solutions 
(\ref{Q3solution1}) and (\ref{Q3solution2}) by any function
$q(\beta_1, \beta_2, \beta_3, l)$ which is $2\pi i$ periodic, 
symmetric, with $q(\beta + i\pi,\beta,\beta_3,l) = {\rm constant}$ and such 
that it reduces to a constant in the $l \rightarrow 0$ limit. 
Such a function may be written, for instance, in terms of 
arbitrary powers of the expression $q_0(\beta_1,\beta_2,\beta_3) = 
\dn\,u_{12}\, \dn\,u_{13}\, \dn\,u_{23}$. 

As a final example, we provide explicit solutions of the recursive
equations (\ref{recursive2}) for $n=4$. In this case, we need to 
find $Q_4$ satisfying
\EQ
Q_4(\beta +i\pi, \beta, \beta_3, \beta_4) = - 4 i\,
\frac{\sn[ 2 (u-u_3)]
+ \sn[2 (u - u_4)]}{
\left[1+\dn[2 (u - u_3)]\right]
\left[1+\dn[2 (u - u_4)]\right]} 
\times Q_2(\beta_3,\beta_4)\,\,\,.
\label{Q4recursive}
\EN
Using some of the identities listed in Appendix A and inserting
the explicit expression for $Q_2$ as given in eq. (\ref{Q2}), the
previous equation can be rewritten as
\EQ
Q_4(\beta +i\pi, \beta, \beta_3, \beta_4) = 
-8i \,
\frac{\sn(2 u - u_3 -u_4) \, \cn^2(u_3-u_4) 
\, \dn(u_3-u_4)}{\left[
\dn(2 u - u_3-u_4)
~+~\dn(u_3-u_4)\right]^2}
\EN
If we now define the functions
\EQ
D_1 ~=~ \prod_{i<j}^4\dn\,u_{{\it ij}}  
\,\,\, ,
\EN
\EQ
D_2 ~=~ \dn(u_{12}+u_{34}) \,  
\dn(u_{12}-u_{34}) \,   
\dn(u_{14}+u_{23}) \,\,\, ,
\EN
it is easy to determine various solutions with no poles on 
the physical strip. Here we list two of them. 

The first solution is
\begin{eqnarray} 
&& Q_4^{(1)}(\beta_1, \beta_2, \beta_3, \beta_4) ~=~
-\frac{4i}{l'} \, \frac{D_1^3 \, D_2}{D_1^3 \, D_2 + (l')^6}
\times \nonumber \\
&&~~~~~~~~
\,
\left( \cn\,u_{12} \cdot
\cn\,u_{34} +
\cn\,u_{13} \cdot
\cn\,u_{24} +
\cn\,u_{14} \cdot
\cn\,u_{23} \right) \,  
\times 
\label{Q4sol1}
\\
&&~~~~~~~~
\left[ 1 +
\frac{\cn(u_{12}+u_{34})}{\dn(u_{12}+u_{34})} \cdot
\frac{\cn(u_{12}-u_{34})}{\dn(u_{12}-u_{34})}  +
\frac{\cn(u_{13}+u_{23})}{\dn(u_{13}+u_{23})} 
\cdot \frac{\cn(u_{13} - u_{23})}{\dn(u_{13} -u_{23})}  ~+~ \right.
\nonumber \\
&&~~~~~~~~~~~~
\left. ~+~
\frac{\cn(u_{14}+u_{24})}{\dn(u_{14}+u_{24})} \cdot 
\frac{\cn(u_{14}-u_{24})}{\dn(u_{14}-u_{24})} \right]
\,\,\,.
\nonumber 
\end{eqnarray}
In the limit $T \rightarrow \infty$ it reduces to the standard
Sinh--Gordon solution \cite{KM} 
\begin{eqnarray}
&& Q_4^{Sh}(\beta_1,\beta_2,\beta_3,\beta_4) = -i \,
\left(
\cosh\frac{\beta_{12}}{2} \,
\cosh\frac{\beta_{34}}{2}  +  
\cosh\frac{\beta_{13}}{2} \,
\cosh\frac{\beta_{24}}{2} + 
\cosh\frac{\beta_{14}}{2} \,
\cosh\frac{\beta_{23}}{2} \right) ~\times \nonumber \\
&&~~~~~
\left[2 + 
\cosh\beta_{12} +
\cosh\beta_{13} + 
\cosh\beta_{14} +  
\cosh\beta_{23} + 
\cosh\beta_{24} +
\cosh\beta_{34} \right]  \label{ShG41} \,\,\,.
\end{eqnarray}
The Sinh--Gordon FF relative to this solution goes asymptotically 
to a constant in each variable $\beta_i$.   

The second solution reads
\begin{eqnarray}
&& Q_4^{(2)}(\beta_1,\beta_2,\beta_3,\beta_4) ~=~ 
 -8i(l')^3 \, \frac{D_1^5 \, D_2^2}{(D_1^3 \, D_2 +
(l')^6)^2} \times 
\nonumber 
\label{Q4sol2}
\\
&&~~~~~~~~~ 
\left[ \frac{\cn(u_{12}+u_{34})}{\dn(u_{12}+u_{34})} \cdot
\frac{\cn(u_{13}
-u_{24})}{\dn(u_{12}-
u_{24})} \cdot \frac{\cn(u_{14}
+u_{23})}{\dn(u_{14}+
u_{23})} ~ +~ \right.\\ 
&&~~~~~~~~~\nonumber \left.
\frac{\cn(u_{12}+u_{34})}{\dn(u_{12}+u_{34})}  
+\frac{\cn(u_{13}-u_{24})}{\dn(u_{12}-u_{24})} +
\frac{\cn(u_{14}+u_{23})}{\dn(u_{14}+u_{23})} \right] \,\,\, ,
\end{eqnarray}
and in the Sinh--Gordon limit it reduces to the second 
standard solution of this model \cite{KM} 
\begin{eqnarray}
&& Q_4^{Sh}(\beta_1,\beta_2,\beta_3,\beta_4) = -2 i \,
\left( 
\cosh\frac{(\beta_{12}+\beta_{34})}{2} \cdot 
\cosh\frac{(\beta_{12}+\beta_{43})}{2} \cdot 
\cosh\frac{(\beta_{14}+\beta_{23})}{2}  
\right. ~ + ~ \nonumber \\
&&~~~~~~ \left.
+ ~ \cosh\frac{(\beta_{12}+\beta_{34})}{2} ~+~ 
\cosh\frac{(\beta_{12}+\beta_{43})}{2} ~+~
\cosh\frac{(\beta_{14}+\beta_{23})}{2} \right) 
\label{ShG42}
\,\,\,.
\end{eqnarray} 
The Sinh--Gordon FF relative to this solution goes asymptotically 
to zero in each variable $\beta_i$.   

Starting from a different expression for $Q_2$, one can easily 
generate other solutions for the four--particle Form Factors which 
however reduce either to (\ref{ShG41}) or (\ref{ShG42}) in the limit 
$l\rightarrow 0$.  
 
Concerning the computation of higher FF, it can be performed along 
the same lines of the previous examples. As shown by the  
first few cases, the periodic nature of the theory will manifest 
itself in the appearance of infinite sets of solutions which all 
collapse to the standard solutions for the Sinh--Gordon model in 
the $l \rightarrow 0$ limit. As far as the modulus $l$ is kept 
away from zero, different solutions for the $n$--particle Form 
Factors are characterized by a different analytic structure.
It would be interesting to understand how to discriminate
among different FF on the basis of a detailed investigation of
the zeros and poles distributions of matrix elements of the various
operators present in the theory. Moreover, it would be highly 
desirable to have a concise and closed formula at least for a 
sequence of Form Factors $F_n(\beta_1,...,\beta_n)$ as in the 
Sinh--Gordon model \cite{KM}, but the determination of this 
formula has presently eluded our attempts.   

Finally, let us briefly discuss some aspects of the ESG model in 
its ultraviolet regime. Among the solutions of the FF equation 
we expect to find those ones which identify with the matrix elements 
of the trace $\Theta(x)$ of the stress--energy tensor. This is 
a $Z_2$ even field, with non--vanishing FF for all $2 n$ numbers 
of external particles. The FF of the field $\Theta(x)$ can be used 
to compute the central charge of the model in its ultraviolet limit 
by means of the $c$-theorem sum rule
\EQ
C(T) \,=\, \int_0^{\infty} d\mu\, c_1(\mu)\,\, ,
\label{variation}
\EN
where $c_1(\mu)$ is given by
\begin{eqnarray}
c_1(\mu)& =&\frac{12}{\mu^3} \sum_{n=1}^{\infty} \frac{1}{(2n)!}
\int\frac{d\beta_1\ldots d\beta_{2n}}{(2\pi)^{2n}}\,
\mid F_{2n}^{\Theta}(\beta_1,\ldots, \beta_{2n})\mid^2 
\label{seriesss} \\
& & \,\,\, \times \,
\delta(\sum_i m\sinh\beta_i)\,\delta(\sum_i m\cosh\beta_i-\mu)\,\,\, .
\nonumber 
\end{eqnarray}   
Assuming the uniform convergence of this series (so that we can  
interchange the sum on the index $n$ with the integral in 
(\ref{variation})), it is immediate to conclude that the central 
charge of the model can take any value in the interval $(0,1)$ 
depending on the period $T$ of the $S$--matrix. In fact, for 
$T \rightarrow \infty$ we recover the sum rule series of the 
Sinh--Gordon model, i.e. $C(\infty) =1$. Viceversa for 
$T \rightarrow 0$ each term of the series goes to zero and 
therefore $C(0) = 0$. The series (\ref{seriesss}) is a continous  
and monotonic decreasing function of $T$ so that by varying 
this parameter we can reach any value in the interval $(0,1)$. 
Therefore, in the ultraviolet limit we have generically a 
non--unitary irrational Conformal Field Theory. As a consequence, 
it seems reasonable that the existence of an infinite dimensional 
space of solutions for the FF of the ESG theory can be traced 
back to the non--unitary irrational nature of the CFT in the 
UV region and to its relative infinite number of primary operators.  

\resection{Conclusions}

In this paper we have analysed certain massive integrable 
$Z_2$ invariant quantum field theories with an infinite 
tower of resonance states. These states are associated 
to an unlimited sequence of poles on the unphysical sheet 
of the $S$--matrix for a fundamental particle $A$ and may 
be regarded as unstable bound states thereof. As a function 
of the energy, the phase--shift of the scattering amplitude 
presents sharp jumps in correspondence to the masses of 
the resonances. The $S$--matrix of these theories has 
a real periodic behaviour in the variable $\beta$. 

The Form Factors of such theories may be computed in 
principle along the same lines of other integrable 
relativistic models, i.e. by solving the monodromy 
and recursive equations coming from the residue 
condition on the kinematical poles. However, new 
features have emerged from the periodic nature of 
the $S$--matrix: on the one hand, there is a very 
severe constraint on the residues of the Form Factors 
$F_n(\beta_1,...\beta_n)$ at their kinematical poles 
$\beta_{ij}= i \pi$, which have to be periodic expressions 
of the remaining rapidity variables $\beta_k$. Conversely, 
an infinite proliferation of solutions of the FF equations 
occurs, due to the lack of very stringent conditions on their 
analytic structure or to the impossibility of enforcing 
a given asymptotic behaviour in each rapidity variable 
$\beta_i$. 

While it is an interesting open problem to develop 
further theoretical criteria to identify the operators 
associated to these solutions, nevertheless the presence 
of an infinite number of them seems compatible with an 
other aspect of these models related to their behavior
in the ultraviolet regime. In fact, the Conformal Field 
Theories which are approached in the ultraviolet limit are 
generally non--unitary and irrational, therefore with an 
infinite number of primary fields. The value of the central 
charge may be fixed by varying the spectrum of the resonances, 
i.e. by changing the period $T$ of the $S$--matrix. 
 
Since these QFT with resonance states may be thought of as 
theories with additional mass scales compared to those given by 
the spectrum of stable particles, it is an interesting 
general problem to determine how these extra mass scales 
affect the different regimes of the theory. From our analysis 
it has emerged that the large distance behaviour of the model 
is hardly affected by them, being essentially ruled by the mass 
gaps of the stable part of the spectrum. On the contrary, the
ultraviolet properties of the theory are deeply influenced by 
the infinite tower of unstable states which, in the ultraviolet 
regime, tend to smooth the short--distance divergencies and 
to decrease their related quantities. As a consequence, 
ultraviolet singularities are in general less severe, 
and the central charge of the underlying Conformal Field 
Theory takes lower (in general irrational) values than in the 
absence of the resonance states. It would be interesting to 
further investigate the ultraviolet properties of these theories 
with resonances by means of the Thermodynamics Bethe Ansatz. 

More generally, it would be highly desirable to understand whether 
these theories allow for a Lagrangian description, and to find 
possible applications to statistical mechanics systems. 
Moreover, the possibility to define other periodic models certainly 
deserves more investigation as well as the possible computation
of FF for the $Z_4$--model of Ref. \cite{Z4Zam}.

\vspace{1cm}
{\bf Acknowledgments} 

This work was done under 
partial support of the EC TMR Programmes FMRX-CT96-0012, and
ERBFMRX-CT96-0045 in which SP is associated to the University of Torino.  

\newpage

\appendix\appsection

In this appendix we collect some useful formulas relative to the 
Jacobian elliptic functions as well as mathematical identities 
used in the text. The reader may consult \cite{Gran} for further 
details. 

The Jacobian elliptic functions are double--periodic functions and 
have two simple poles and two simple zeros in a period parallelogram. 
Let 
\EQ
\ka(l) = \int_0^{\frac{\pi}{2}} \frac{d\alpha}
{\sqrt{1 - l^2 \sin^2\alpha}} \,\,\, ,
\label{completeell}
\EN 
be the complete elliptic integral of modulus $l$ and $\ka'(l)= 
\ka(l')$ the complete elliptic integral of the complementary 
modulus $l'$, with $l^2 + l'^2 =1$. Setting $q = \exp\left[-\pi 
\ka'/\ka\right]$, the definition of the Jacobian elliptic 
functions $\sn {\it u}$, $\cn {\it u}$ and $\dn {\it u}$ is given by 
\[
\hspace{-5mm}
\sn {\it u} = \frac{2 q^{\frac{1}{4}}}{\sqrt {\it l}} \,\sin
\frac{\pi {\it u}}{2 \ka} \,
\prod_{n=1}^{\infty} 
\frac
{1 - 2 q^{2 n} \cos\frac{\pi {\it u}}{\ka} + q^{4 n}}
{1 - 2 q^{2 n-1} \cos\frac{\pi {\it u}}{\ka} + q^{4 n-2}} \,\,\, ,
\]
\EQ
\,\,\,\,\cn {\it u} = \frac{2 \sqrt{{\it l}'} \,
q^{\frac{1}{4}}}{\sqrt {\it l}} \,
\cos\frac{\pi {\it u}}{2 \ka} \,
\prod_{n=1}^{\infty} 
\frac
{1 + 2 q^{2 n} \cos\frac{\pi {\it u}}{\ka} + q^{4 n}}
{1 - 2 q^{2 n-1} \cos\frac{\pi {\it u}}{\ka} + q^{4 n-2}} \,\,\,,
\EN
\[\hspace{-2cm}
\dn {\it u} = {\sqrt {{\it l}'}} \,
\prod_{n=1}^{\infty} 
\frac
{1 + 2 q^{2 n -1} \cos\frac{\pi {\it u}}{\ka} + q^{4 n - 2}}
{1 - 2 q^{2 n-1} \cos\frac{\pi {\it u}}{\ka} + q^{4 n-2}} \,\,\,.
\]
The parity properties of the Jacobian elliptic functions are
\EQ
\cn(-{\it u})~=~\cn{\it u} \quad ; \quad
\sn(-{\it u})~=~-\sn{\it u} \quad ; \quad 
\dn(-{\it u})~=~\dn{\it u} \,\,\,,
\EN
whereas their value in $u=0$ is
\EQ
\cn(0) ~=~ \dn( 0) ~=~1 \quad ; \quad \sn(0)~=~0 \,\,\,.
\EN
The period, zeros and poles of these functions are 
summarised in the following table  
$$ 
\begin{array}{|c|c|c|c|}
 \hline
& & & \\
{\rm Function} 
& {\rm Periods}
& {\rm Zeros}
& {\rm Poles}
\\
%   &        &       &  \\
\hline
   &        &       &  \\
\sn {\it u}& 4m\ka + 2n \ka'i  & 2m\ka + 2n\ka'i & 2m\ka + (2n+1)\ka'i \\
%   &        &       &  \\
\hline
   &        &       &  \\
\cn {\it u} & 4m\ka + 2n(\ka +\ka'i) & (2m+1)\ka + 2n\ka'i & 2m\ka +
(2n+1)\ka'i
 \\ %   &        &       &  \\
\hline
   &        &       &  \\
\dn {\it u} &  2m\ka + 4n\ka'i &  (2m+1)\ka +(2n+1)\ka'i & 2m\ka +(2n+1)\ka'i
 \\ %   &        &       &  \\ 
\hline 
\end{array}
$$

\vspace{3mm}

\begin{center}
{\bf Table 1} 
\end{center}
The change of argument of these functions are ruled by the table 
$$ 
\begin{array}{|c|c|c|c|c|}
 \hline
 & &  & & \\
{\it u}^* = {\it u} + \ka &
{\it u} + i \ka' & 
{\it u} + 2 \ka & 
{\it u} + 2 i \ka' &
{\it u} + 2 \ka + 2 i \ka' 
\\
   &        &         & &  \\
\hline
   &        &         & & \\
\sn {\it u}^* = \frac{{\cn} {\it u}}{{\dn} {\it u}} & \frac{1}{l\, 
\sn {\it u}} &
 -\sn {\it u} & 
\sn {\it u} & - \sn {\it u} \\
   &        &         & & \\
\hline
   &        &         & & \\
\cn {\it u}^* = -{\it l}' \,\frac{\sn {\it u}}{\dn {\it u}} &
-\frac{i}{l}\, \frac{\dn {\it u}}{\sn {\it u}} 
 & -\cn {\it u} & - \cn {\it u}& 
\cn {\it u} \\
   &        &         & & \\
\hline
   &        &         & & \\
\dn {\it u}^* = {\it l}' \,\frac{1}{\dn {\it u}} & - i 
\frac{\cn {\it u}}{\sn {\it u}} & 
\dn {\it u} & - \dn {\it u}& - \dn {\it u} \\
 &   &        &       & \\ 
\hline 
\end{array}
$$

\vspace{3mm}

\begin{center}
{\bf Table 2} 
\end{center}
We list here all the indentities satisfied by the Jacobian elliptic
functions used in the course of our calculations
\EQ
\begin{array}{ll}
& \cn^2 {\it u} ~+~ \sn^2 {\it u} ~=~ 1 \,\,\,;\nonumber \\
& \dn^2 {\it u} ~+~ {\it l}^2 \,\sn^2 {\it u} ~=~ 1 \,\,\,;
\end{array}
\EN
\[
\sn\left({\it u}\pm {\it v}\right) ~=~ \frac{\sn {\it u}\, \cn {\it v} 
\,\dn {\it v}\, \pm \sn{\it v} \, \cn{\it u} \, 
\dn{\it u}}{1 - {\it l}^2 \,\sn^2{\it u} \,\sn^2{\it v}} \,\,\,;
\]
\EQ
\cn\left({\it u}\pm {\it v}\right) ~=~ \frac{\cn{\it u} \,\cn{\it v} 
\mp \sn{\it u} \,\sn{\it v} \,\dn{\it u} \,\dn{\it v}}{1 - {\it l}^2
\,\sn^2{\it u} 
\,\sn^2{\it v}} \,\,\,;
\label{addition}
\EN
\[
\,\,\,\dn\left({\it u}\pm {\it v}\right) ~=~ \frac{\dn{\it u} \,
\dn{\it v} \mp 
{\it l}^2 \,\sn{\it u} \,\sn{\it v} \,\cn{\it u} \,
\cn{\it v}}{1- {\it l}^2\, \sn^2{\it u}  \,\sn^2{\it v}} \,\,\,;
\]
\EQ
\sn({\it u} +{\it v}) \, \sn({\it u}-{\it v}) ~=~ \frac{\sn^2{\it u} -
\sn^2{\it v}}{1-{\it l}^2 \,\sn^2{\it u} \, \sn^2{\it v}}
\EN
\EQ
\cn({\it u} +{\it v}) \, \cn({\it u}-{\it v}) ~=~ \frac{\cn^2{\it v} -
\sn^2{\it u}\, \dn^2{\it v}}{1-{\it l}^2 \,\sn^2{\it u} \, \sn^2{\it v}}
\EN
\EQ
\dn({\it u} +{\it v}) \, \dn({\it u}-{\it v}) ~=~ \frac{\dn^2{\it v} -
{\it l}^2 \sn^2{\it u} \, 
\cn^2{\it v}}{1-{\it l}^2 \,\sn^2{\it u} \, \sn^2{\it v}}
\EN
\EQ
\frac{\sn {\it u} \, \cn {\it u}}{\dn {\it u}} ~=~ 
\frac{\sn(2{\it u})}{1~+~\dn(2{\it u})} \,\,\,.
\label{1+dn2}
\EN
\EQ
\sn^2 {\it u} ~+~ \frac{\cn^2 {\it u}}{\dn^2 {\it u}} ~=~
\frac{2}{1~+~\dn(2{\it u})}
\EN  
When the module $l$ goes to zero, we have the approximations 
\[
\sn {\it u} \simeq \sin {\it u} - 
\frac{1}{4} \,{\it l}^2 ({\it u} - 
\sin {\it u} \cos {\it u}) \cos {\it u} \,\,\,;
\]
\EQ
\cn {\it u} \simeq \cos {\it u} + 
\frac{1}{4} {\it l}^2 \,({\it u} - 
\sin {\it u} \cos {\it u}) \sin {\it u} \,\,\,;
\EN
\[
\hspace{-33mm}
\dn {\it u} \simeq 1 - \frac{1}{2} \,{\it l}^2 \sin^2 {\it u} \,\,\,.
\]

Finally, we also report some useful identities concerning 
infinite sums of trigonometric functions. These identities are
useful for calculations in Sections 3 and 5. 
\EQ
\sum_{n=1}^{\infty} \frac{\sin(2 n -1)x }{2 n -1} 
= \frac{\pi}{4} \,\,\,, 
\label{usefulidentity}
\EN 
\EQ
\sum_{n=1}^{\infty} \frac{\sin(nx) }{n} 
= \frac{\pi-x}{2} \,\,\,.
\label{usefulidentity2}
\EN 
As a consequence of the previous identity we also have
\EQ
\sum_{n=1}^{\infty} \frac{(-1)^{n-1}}{n} \sin n x = \frac{x}{2}
\,\,\, .
\label{usefulidentity3}
\EN 
Other useful identities are
\EQ
\sum_{n=1}^{\infty} \frac{1}{2n-1} \, q^{2n-1} \, \cos{(2n-1)x} ~=~ 
\frac{1}{4} \ln \frac{1+2q \cos{x} +q^2}{1 -2q\cos{x} +q^2}
\label{usefulidentity4} \,\,\,,
\EN
\EQ
\sum_{n=1}^{\infty}  \frac{q^n}{n} \cos(nx) ~=~ 
-\frac{1}{2} \, \ln \left(1-2q\cos x + q^2 \right) 
\label{usefulidentity12} \,\,\,,
\EN
\EQ
\left|\frac{\Gamma(x)}{\Gamma(x + i y)}\right| ~=~
\prod_{k=0}^{\infty} \left(1 + \frac{y^2}{(x+k)^2}\right) 
\,\,\,.
\label{usefulidentity13} 
\EN

\newpage 

\appsection

In this appendix we present the calculations which leads to 
the exponential representation (\ref{firstexp}). 

Due to the periodicity of the S--matrix for real values of the 
rapidity, we can always look for its Fourier series expansion 
with period $T$ of the form 
\EQ
S(\beta,a) = - \exp\left[\sum_{n=-\infty}^{+\infty} \, 
a_n \, e^{\frac{2 i \pi}{T} n \beta} \right] \,\,\,.
\label{fourier}
\EN 
In order to determine the coefficients $a_n$, let us equating the 
logarithmic derivative of the above expression with the one 
coming from the $S$--matrix (\ref{2ellipS}) 
\EQ
\sum_{n=-\infty}^{+\infty} a_n \, n \, e^{\frac{2 i\pi}{T} n \beta} = 
-\frac{2\ka'}{\pi} \, E(\beta) \,\,\, ,
\label{fou}
\EN 
where 
\EQ
E(\beta) = \sn\left(2\ka {\it a}\right) \, 
\frac{\sn'\left(\frac{2 \ka i \beta}{\pi}\right)}
{\left(\sn\left(\frac{2 \ka i \beta}{\pi}\right)
+ \sn\left(2\ka {\it a}\right)\right)\, 
\left(\sn\left( \frac{2 i \ka \beta}{\pi}\right) 
- \sn\left(2 \ka {\it a}\right)\right)} 
\label{left}
\EN
and $\sn'({\it x})$ denotes the derivative of the function 
$\sn({\it x})$. 
The coefficients $a_n$ for $n \not= 0$ are then given by
\EQ
n \, a_n ~=~ -\frac{2 \ka'}{\pi T} \, \int_{-\frac{T}{2}}^{\frac{T}{2}}
\, dt \, \, e^{-2i\frac{\pi}{T} n t} \, E(t)
\EN
Let us consider the complex loop integral 
\EQ
\oint_C \, dt \, e^{-\frac{2 i\pi}{T} n t} \, E(t)
\label{loop}
\EN
on a closed path $C$ which runs along the boundary of half the 
fundamental domain, i.e. $-\frac{T}{2} < {\rm Re} t < \frac{T}{2}$, 
$0 < {\rm Im} t < i\pi$. Due to the real periodicity of the function 
$E(t)$ on the complex $t$--plane, it is easy to see that the contributions
from the vertical lines of the path cancel each others, while the 
integral on the   
${\rm Im}t = i\pi$ line can be rewritten as an integral on the real axis by 
exploiting the property $E(t + i\pi) = -E(t)$. Therefore, for the 
coefficients $a_n$ we have 
\EQ
n \, a_n \, \left( 1 ~+~ e^{2n\frac{\pi^2}{T}} \right) ~=~
-\frac{2 \ka'}{\pi T} \, \oint_C dt \, e^{-\frac{2 i\pi}{T} n t} \, 
E(t)
\EN
Inside the domain of integration the function $E(t)$ 
has two simple poles at $t =i\pi a, \, i\pi(1-a)$. 
Computing the corresponding residues we obtain
\EQ
n \, a_n ~=~ \frac{ \cosh\left(\frac{n \pi^2}{T} 
(1-2a)\right)}{\cosh\left(\frac{n\pi^2}{T}\right)} 
\,\,\,\,\, ,\qquad n \not= 0 
\EN
Inserting this result in eq.\,(\ref{fourier}) and exploiting 
the parity properties of the expression under $n \rightarrow 
-n$, we can write 
\EQ
S(\beta,a) = - \exp\left[
2 i \sum_{n=1}^{\infty} \frac{1}{n} \,\frac{ 
\cosh\left(\frac{n \pi^2 (1 - 2 a)}{T} \right)}
{\cosh\left(\frac{n \pi^2}{T}\right)}  \,
\sin\left(\frac{2 n \pi}{T} \beta\right) ~+~ a_0 \right] \,\,\,,
\label{phase1}
\EN
where $a_0$ is the contribution of the zero--mode, which is given 
by 
\EQ
a_0 = \lim_{n\rightarrow 0} \frac{i}{n} 
\frac{ \cosh\left(\frac{n \pi^2 (1 - 2 a)}{T} \right)}
{\cosh\left(\frac{n \pi^2}{T}\right)}  \,
\sin\left(\frac{2n \pi\beta}{T}\right) = \frac{2 i \pi \beta}{T} \,\,\,.
\label{zeromode}
\EN 
The zero mode $a_0$ ensures in particular that in the limit 
$a \rightarrow 0$ we have $S(\beta,a) \rightarrow 1$. Now, using  
the identity (\ref{usefulidentity3}) 
to rewrite the zero mode as a series expansion, we finally
obtain the expression 
\EQ 
S(\beta) = - \exp\left[2 i \,
\sum_{n=1}^{+\infty} \frac{1}{n} \, \left(
\frac{\cosh\left(\frac{n
\pi^2 (1 - 2 a)}{T}\right)} {\cosh\left(\frac{n \pi^2}{T}\right)} 
- (-1)^n \right) 
\, \sin\left(\frac{2 n \pi}{T} \beta\right)
\right] \,\,\, ,
\EN 
given in the text, eq.\,(\ref{firstexp}). 

In order to obtain the second expression (\ref{secondexp}) for 
the $S$--matrix, it is initially convenient to split the 
zero--mode contribution into two series on the odd and even  
numbers, i.e. 
\begin{eqnarray}
& & 2 i \sum_{n=1}^{+\infty} \frac{(-1)^{n-1}}{n} 
\sin\left(\frac{2 n \pi \beta}{T}\right) =  
\label{intermediate}
\nonumber \\
& &\,\,\, =
2 i \sum_{n=1}^{+\infty} \frac{1}{2 n - 1} 
\sin\left(\frac{2 (2 n -1) \pi \beta}{T}\right) - 
2 i \sum_{n=1}^{+\infty} \frac{1}{2 n} 
\sin\left(\frac{4 n \pi \beta}{T}\right) \,\,\,.
\nonumber 
\end{eqnarray}
By using now the representation (\ref{Ising})
\[ 
-1 = \exp\left[-4 i
\,\sum_{n=1}^{+\infty} \frac{1}{2 n -1} 
\sin\left(\frac{2\pi (2 n -1) \beta}{T}\right) 
\right] \,\,\,, 
\] 
and adding the term in the exponential of this expression 
to the zero mode, as a result we have that the series 
of the zero mode on the odd numbers changes its sign in front so 
that we end up with the expression  
\EQ
\exp\left[-2 i \sum_{n=1}^{+\infty} \frac{1}{n} 
\,\sin\left(\frac{2 n \pi \beta}{T}\right) 
\right] \,\,\,, 
\EN 
By taking into account the previous Fourier expansion we have then 
\begin{eqnarray}
S(\beta,a) & = & \exp\left[2 i \,
\sum_{n=1}^{+\infty} \frac{1}{n} \, \left(
\frac{\cosh\left(\frac{n
\pi^2 (1 - 2 a)}{T}\right)} {\cosh\left(\frac{n \pi^2}{T}\right)} 
- 1 \right) 
\, \sin\left(\frac{2 n \pi}{T} \beta\right)
\right] \,= \\
& =& \exp\left[-4 i \, \sum_{n=1}^{+\infty} \frac{1}{n} \, \frac{
\sinh\left(\frac{n a \pi^2}{T}\right) \, \sinh\left(\frac{n (1-a)
\pi^2}{T}\right) } {\cosh\left(\frac{n \pi^2}{T}\right)} \,
\sin\left(\frac{2 n \pi}{T} \beta\right)\right] \,\,\,. 
\nonumber
\end{eqnarray}
i.e. the expression (\ref{secondexp}) in the text.

\newpage

\appsection

In this appendix we report the detailed calculation for the
ESG minimal Form Factor as given in eq. (\ref{ESGmin2}).
Let us consider the minimal solution to the monodromy equations
for the ESG model as given in eq. (\ref{ESGmin}). By exploiting 
the arbitrariness of an overall constant in front of $F_{\rm min}$ 
we can replace $-2\sin^2\left(n \frac{\pi}{T}\hat{\beta} \right)$ 
in eq. (\ref{ESGmin}) with $\cos\left(\frac{2n\pi}{T}\hat{\beta} 
\right)$ and take as starting point the following expression
\EQ
F_{\rm min}(\beta) ~=~ \exp \left[2 \sum_{n=1}^{\infty} \frac{1}{n}
\frac{\sinh\left(\frac{n{\it a} \pi^2}{T}\right) \sinh\left(
\frac{n(1-{\it a}) \pi^2}{T}\right)}{\cosh\left(\frac{n \pi^2}{T}\right)  
\sinh\left(\frac{2n \pi^2}{T}\right)} \cos \left(\frac{2n \pi}{T} 
\hat{\beta} \right)\right] \,\,\,.
\label{ESGmin3}
\EN
Let us first concentrate on the $n$ coefficient in the previous expansion
\EQ
c_n \equiv \frac{\sinh\left(\frac{n{\it a} \pi^2}{T}\right) \sinh\left(
\frac{n(1-{\it a}) \pi^2}{T}\right)}{\cosh\left(\frac{n \pi^2}{T}\right)  
\sinh\left(\frac{2n \pi^2}{T}\right)} \,\,\,.
\label{coeff}
\EN
If we use the exponential representation for hyperbolic functions and the
series expansions 
\EQ
(\cosh x)^{-1} ~=~  -2 \, \sum_{k=1}^{\infty} (-1)^k \, 
e^{-(2k-1)x} \,\,\, ,
\label{usefulidentity10}
\EN
\EQ
(\sinh x)^{-1} ~=~  2 \, \sum_{k=1}^{\infty} 
e^{-(2k-1)x} \,\,\, ,
\label{usefulidentity11}
\EN
it can be rewritten as
\EQ
c_n =  
\sum_{k=1}^{\infty} \sum_{p=1}^{\infty} (-1)^{k+1}  e^{-2n[(2k-1)
\frac{\pi^2}{2T} +(2p-1) \frac{\pi^2}{T}]} 
\EN
\[
\hspace{3cm}
\times \,\left( e^{-n\frac{\pi^2}{T}
(1-2a)} + e^{n\frac{\pi^2}{T} (1-2a)} - e^{n \frac{\pi^2}{T}}
- e^{-n\frac{\pi^2}{T}} \right) \,\,\,. 
\]
Now, splitting the $k$--sum into a sum on even and odd integers 
and using the general identity
\[
\sum_{k=1}^{\infty} \sum_{p=1}^{\infty} f(k+p) ~=~  
\sum_{k=2}^{\infty} (k-1) \, f(k) \,\,\, ,
\]
we end up with
\EQ
c_n = -     
\sum_{k=2}^{\infty} (k-1) \, \left[ 
q_1^n(k) + q_2^n(k) + q_3^n(k) - q_4^n(k)
-q_5^n(k) - q_6^n(k)\right] \,\,\, ,
\label{cnnn}
\EN 
where we have defined   
\EQ
\begin{array}{ll}
q_1(k) = e^{-2(2k - 1)\frac{\pi^2}{T} } \qquad ; & 
\quad
q_2(k) = e^{-2(2k-2-a)\frac{\pi^2}{T}} \quad ; 
\\
q_3(k) = e^{-2(2k-3+a)\frac{\pi^2}{T}} \qquad ; & 
\quad 
q_4(k) =  e^{-2(2k-1-a)\frac{\pi^2}{T}} \quad ; 
\\
q_5(k) =  e^{-2(2k-2+a)\frac{\pi^2}{T}} \quad ; & 
\quad
q_6(k) = e^{-2(2k-3)\frac{\pi^2}{T}} \quad .
\end{array}
\label{qfactors}
\EN
By plugging the expression (\ref{cnnn}) of $c_n$ into eq.(\ref{ESGmin3}) 
and performing explicitly the sum on $n$ by means of the identity
(\ref{usefulidentity12}), we arrive to the final expression reported 
in the text (eq. (\ref{ESGmin2}))
\EQ
F_{\rm min}(\beta) ~=~
\prod_{k=2}^{\infty} \left( W(\beta,k) \right)^{k-1} \,\,\, ,
\label{ESGmin4}
\EN 
where 
\[
W(\beta,k) = \frac{
\left(1-2q_6 \cos\left(2\frac{\pi}{T} \hat{\beta}\right) +q_6^2\right)
\left(1-2q_5 \cos\left(2\frac{\pi}{T} \hat{\beta}\right) +q_5^2\right)
\left(1-2q_4 \cos\left(2\frac{\pi}{T} \hat{\beta}\right) +q_4^2\right)}
{\left(1-2q_1 \cos\left(2\frac{\pi}{T} \hat{\beta}\right) +q_1^2\right)
\left(1-2q_2 \cos\left(2\frac{\pi}{T} \hat{\beta}\right) +q_2^2\right)
\left(1-2q_3 \cos\left(2\frac{\pi}{T} \hat{\beta}\right) +q_3^2\right)}
\]
It is now easy to determine the location of zeros and poles of this
function. They are simply given by the solutions of the generic
equation $1-2q \cos x +q^2 =0$, for the numerator and denominator
factors, respectively. Rewriting this equation as 
\EQ
\left( 1-q e^{ix}\right)\left( 1-q e^{-ix}\right)~=~ 0 \,\,\, ,
\EN
the general solution has the form
\EQ
x ~=~ \pm i \ln q ~+~2n\pi \,\quad, \quad n \in {\bf Z} \,\,\,.
\EN
Applying this result to the factors in eq. (\ref{ESGmin4}) and
taking in account the explicit expressions for the $q_i$'s,
we obtain the sets of poles and zeros (\ref{ESGpoles}), (\ref{ESGzeros}) 
reported in the text.

\newpage

\hs 

{\bf Figure Captions}

\vspace{5mm}

\begin{description}
\item [Figure 1.a]. Tiling of the $\beta$--plane in terms of 
the periodic domains of the $S$--matrix. 
\item [Figure 1.b]. Twists of the elastic branch--cuts 
of the $S$--matrix in the $s$--plane.
\item [Figure 2]. Analytic structure of the elliptic $S$--matrix 
in its fundamental domain. The circles are the zeros of the function 
and the black circles its poles. 
\item [Figure 3]. Phase--shift (in units of $\pi$) of the 
Elliptic Sinh--Gordon $S$--matrix 
for particular values of the parameters ($l=0.25$ and $a=0.3$). 
\item [Figure 4]. Phase--shift (in units of $\pi$) of the Elliptic 
Ising Model with the period relative to $l=0.25$. 
\item [Figure 5]. Poles and zeros in the $\beta$--plane for 
$a = \delta - i \frac{T}{2\pi}$. They are split by $2 \delta$ 
and annihilate each other for $\delta \rightarrow 0$. In this 
figure the rectangle delimitated by dashed lines is taken as 
the fundamental domain.   
\item [Figure 6]. Diagram associated to the pole singularity 
of the $S$--matrix near the resonance $R_n$. 
\item [Figure 7]. Plots of the two--point correlation function 
of the field $\Theta(x)$ versus 
${\cal M} R$. The full line is relative to the usual thermal 
Ising model ($l = 0$), the short--dashed line to the value 
$l = 0.25$ and the long--dashed line to $l = 0.75$. 
\item [Figure 8]. Central charge $c$ versus the 
modulus $l$ for the Elliptic Ising model obtained by the 
$c$--theorem sum--rule.  
\item [Figure 9]. Anomalous dimension of $\sigma(x)$ and $\mu(x)$ 
versus the modulus $l$ for the Elliptic Ising model, 
obtained by the $\eta$ sum rule. 
\item [Figure 10]. Analytic structure of $F_{\rm min}(\beta)$ of the 
ESG model along the imaginary direction placed at $\beta = n T$. 
The circles are the zeros of the function whereas the black circles 
represent  its poles. 
\item [Figure 11]. Graphical multiplication of $F_{\rm min}(\beta)$
by $F_{\rm min}(\beta + i \pi)$. 
\end{description}

\newpage 

\pagestyle{empty}

\begin{figure}
%\null\vskip 3cm
\hspace{1cm} \psfig{figure=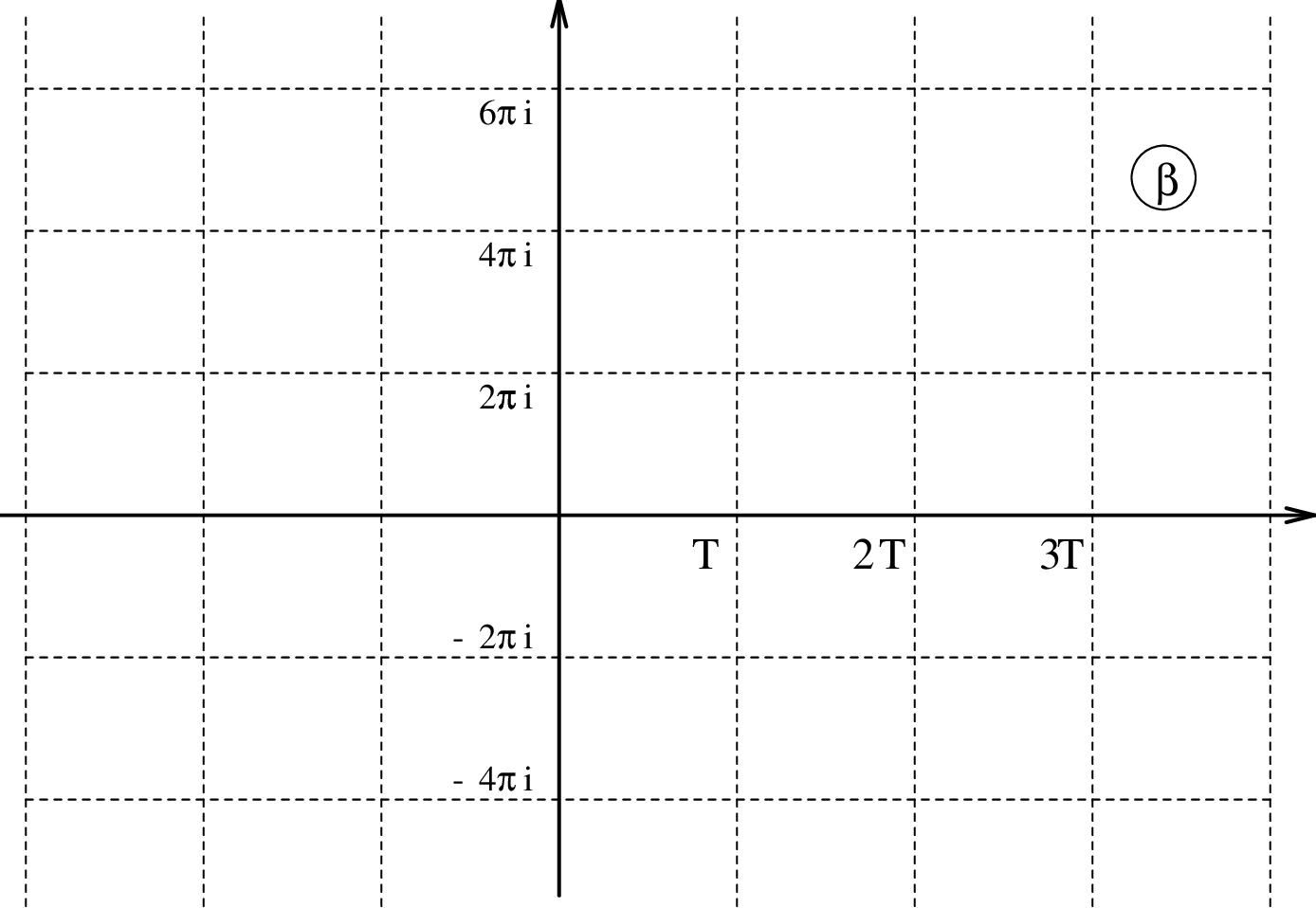,height=10cm,width=15cm}
\vspace{1cm}
\begin{center}   
{\bf \large{Figure 1.a}}
\end{center}
\end{figure}

\newpage

\pagestyle{empty}

\begin{figure}
%\null\vskip 3cm
\hspace{1cm} \psfig{figure=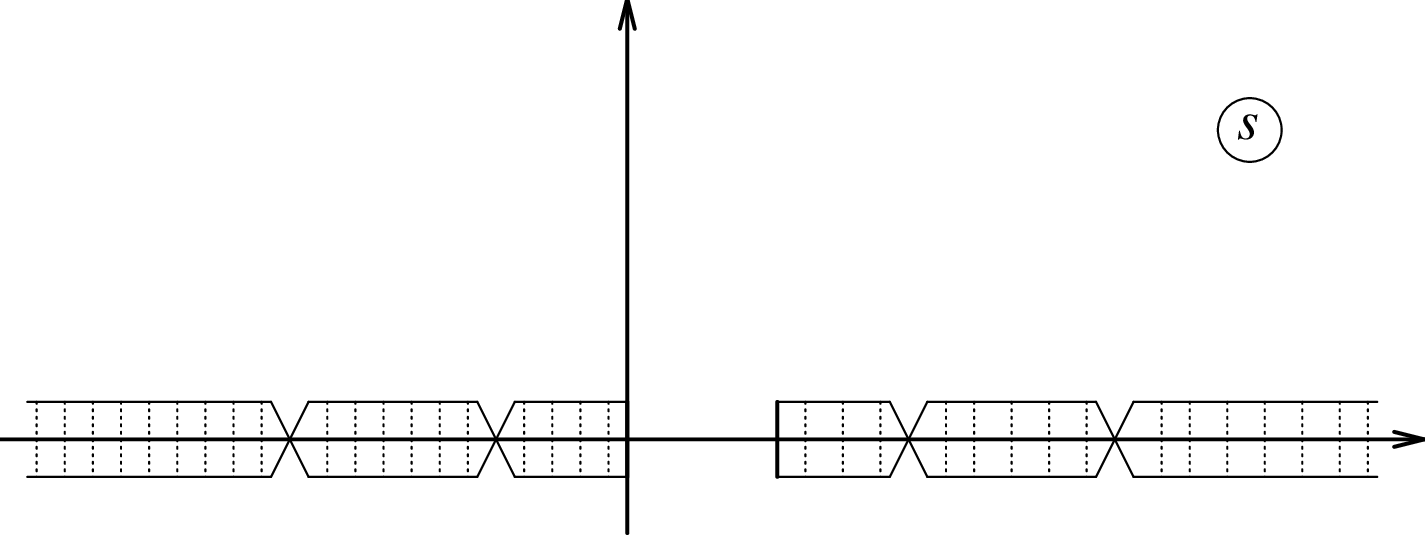,height=8cm,width=15cm}
\vspace{1cm}
\begin{center}   
{\bf \large{Figure 1.b}}
\end{center}
\end{figure}

\newpage

\pagestyle{empty}

\begin{figure}
%\null\vskip -3cm
\hspace{1cm} \psfig{figure=analytics.ps,height=10cm,width=15cm}
%\vspace{1cm}
\begin{center}   
\hspace{15mm}{\bf \large{Figure 2}}
\end{center}
\end{figure}

\newpage

\pagestyle{empty}

\begin{figure}
%\null\vskip -3cm
\hspace{1cm} \psfig{figure=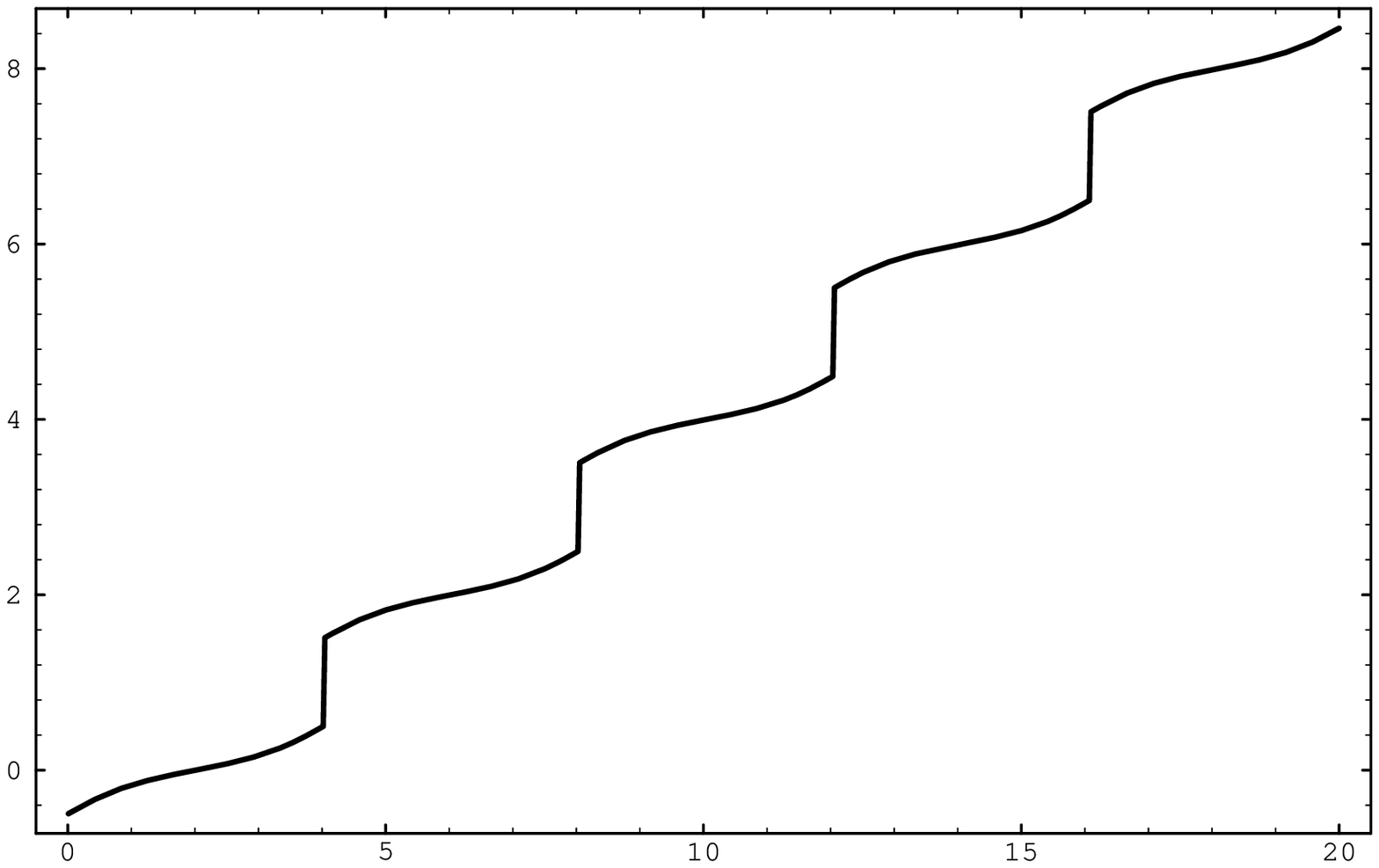,height=25cm,width=15cm}

\vspace{-3cm}

\begin{center}   
\hspace{10mm}{\bf \large{Figure 3}}
\end{center}
\end{figure}

\newpage

\pagestyle{empty}

\begin{figure}
%\null\vskip -3cm
\hspace{1cm} \psfig{figure=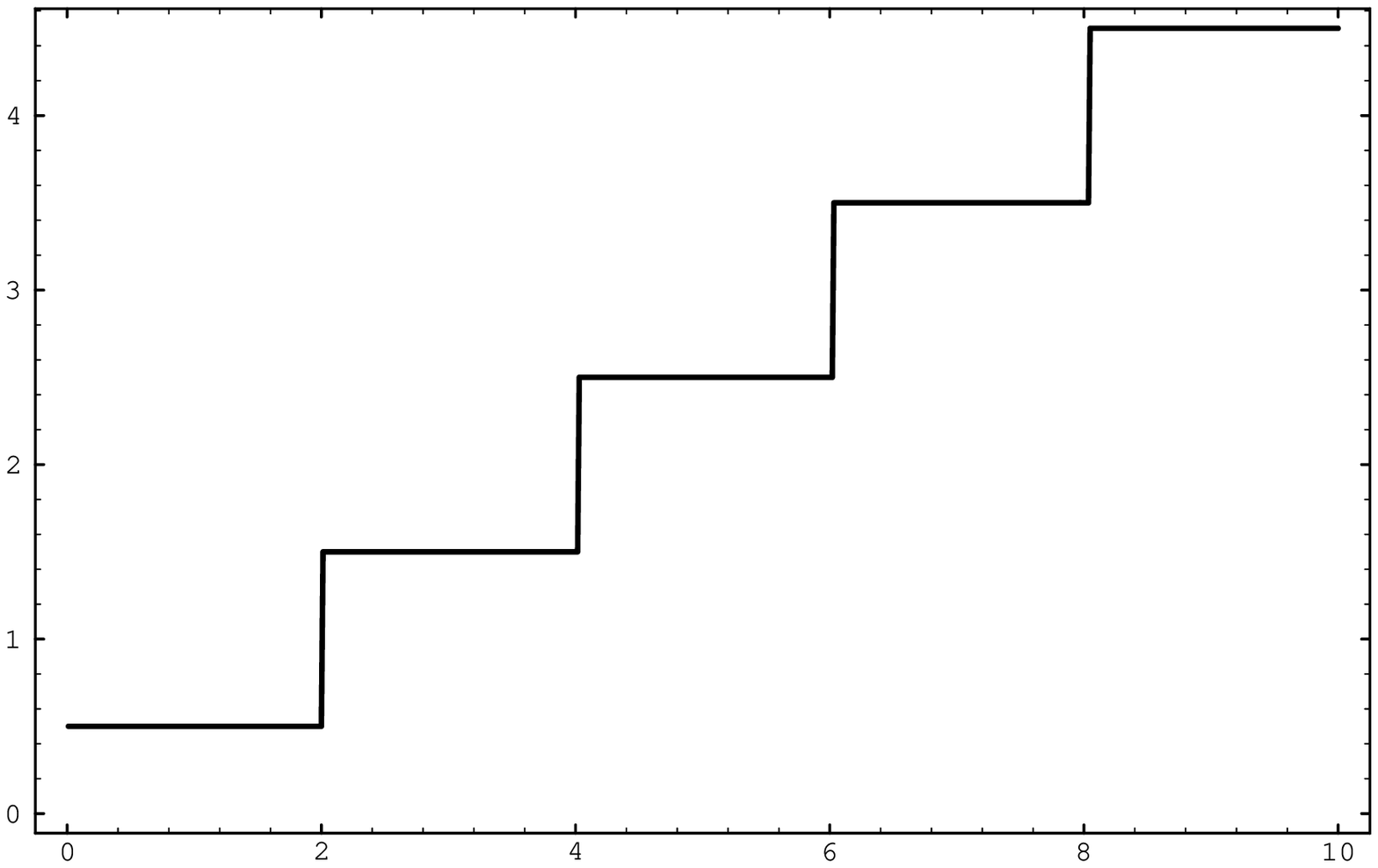,height=25cm,width=15cm}

\vspace{-3cm}

\begin{center}
\hspace{10mm}   
{\bf \large{Figure 4}}
\end{center}
\end{figure}

\newpage

\pagestyle{empty}

\begin{figure}
%\null\vskip -3cm
\psfig{figure=polezeroising.ps,height=10cm,width=15cm,angle=90}

\vspace{1cm}

\begin{center}
\hspace{10mm}   
{\bf \large{Figure 5}}
\end{center}
\end{figure}

\newpage

\pagestyle{empty}

\begin{figure}
%\null\vskip -3cm
\psfig{figure=residue.ps,height=10cm,width=15cm}

\vspace{1cm}

\begin{center}
\hspace{10mm}   
{\bf \large{Figure 6}}
\end{center}
\end{figure}

\newpage

\pagestyle{empty}

\begin{figure}
%\null\vskip -3cm
\hspace{1cm} \psfig{figure=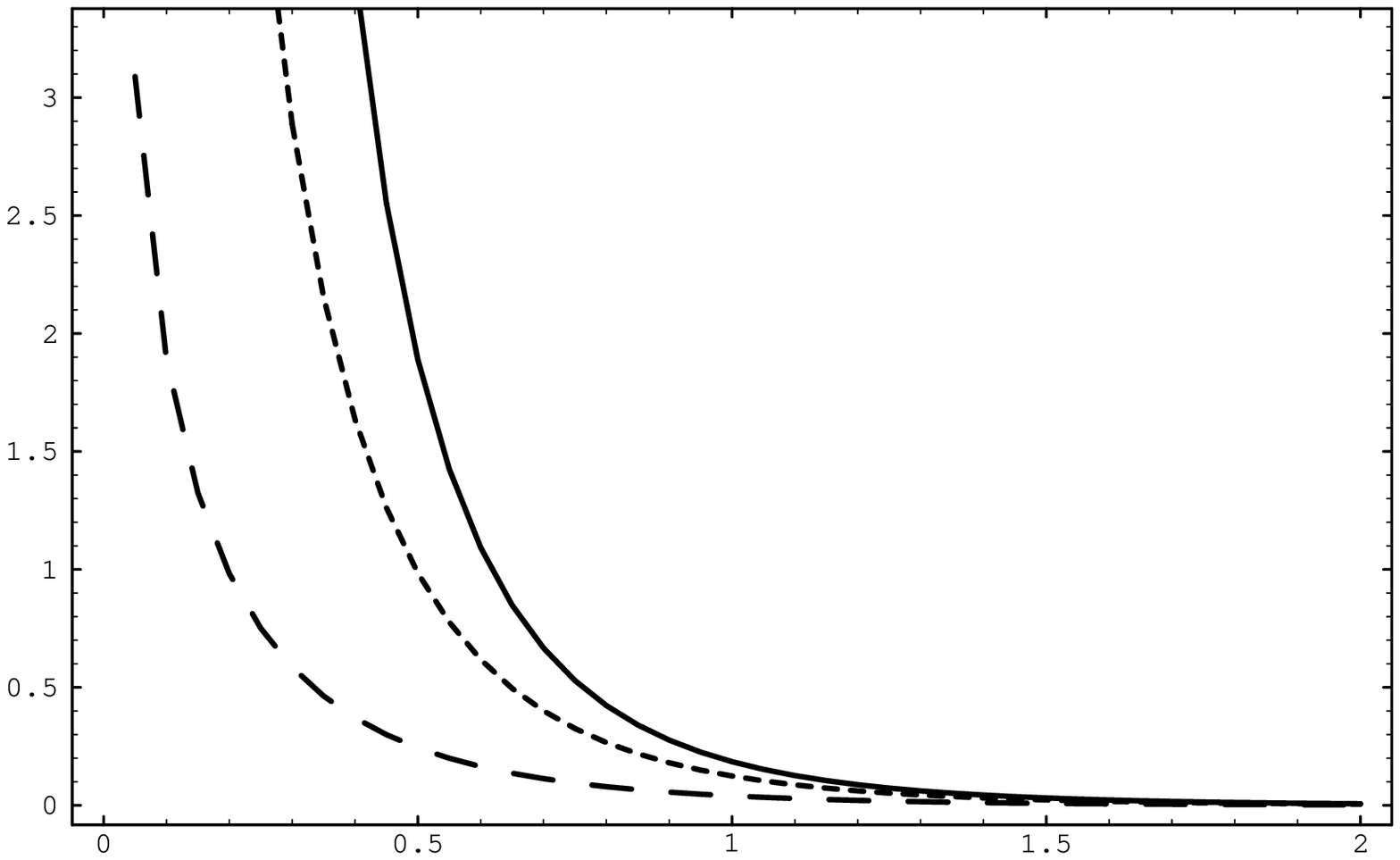,height=25cm,width=15cm}

\vspace{-3cm}

\begin{center}
\hspace{10mm}   
{\bf \large{Figure 7}}
\end{center}
\end{figure}

\newpage

\pagestyle{empty}

\begin{figure}
%\null\vskip -3cm
\hspace{1cm} \psfig{figure=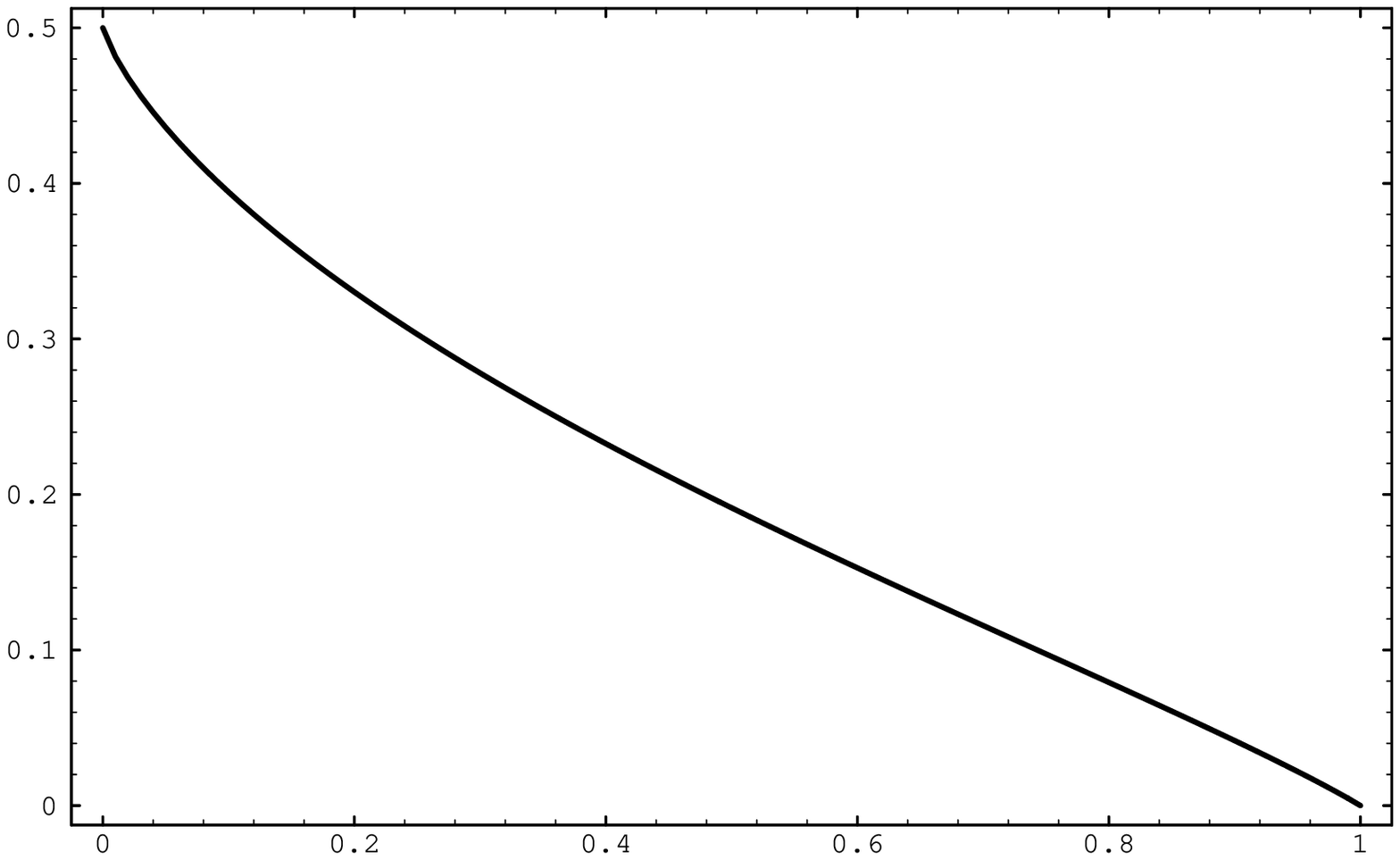,height=25cm,width=15cm}

\vspace{-3cm}

\begin{center}
\hspace{10mm}   
{\bf \large{Figure 8}}
\end{center}
\end{figure}

\newpage

\pagestyle{empty}

\begin{figure}
%\null\vskip -3cm
\hspace{1cm} \psfig{figure=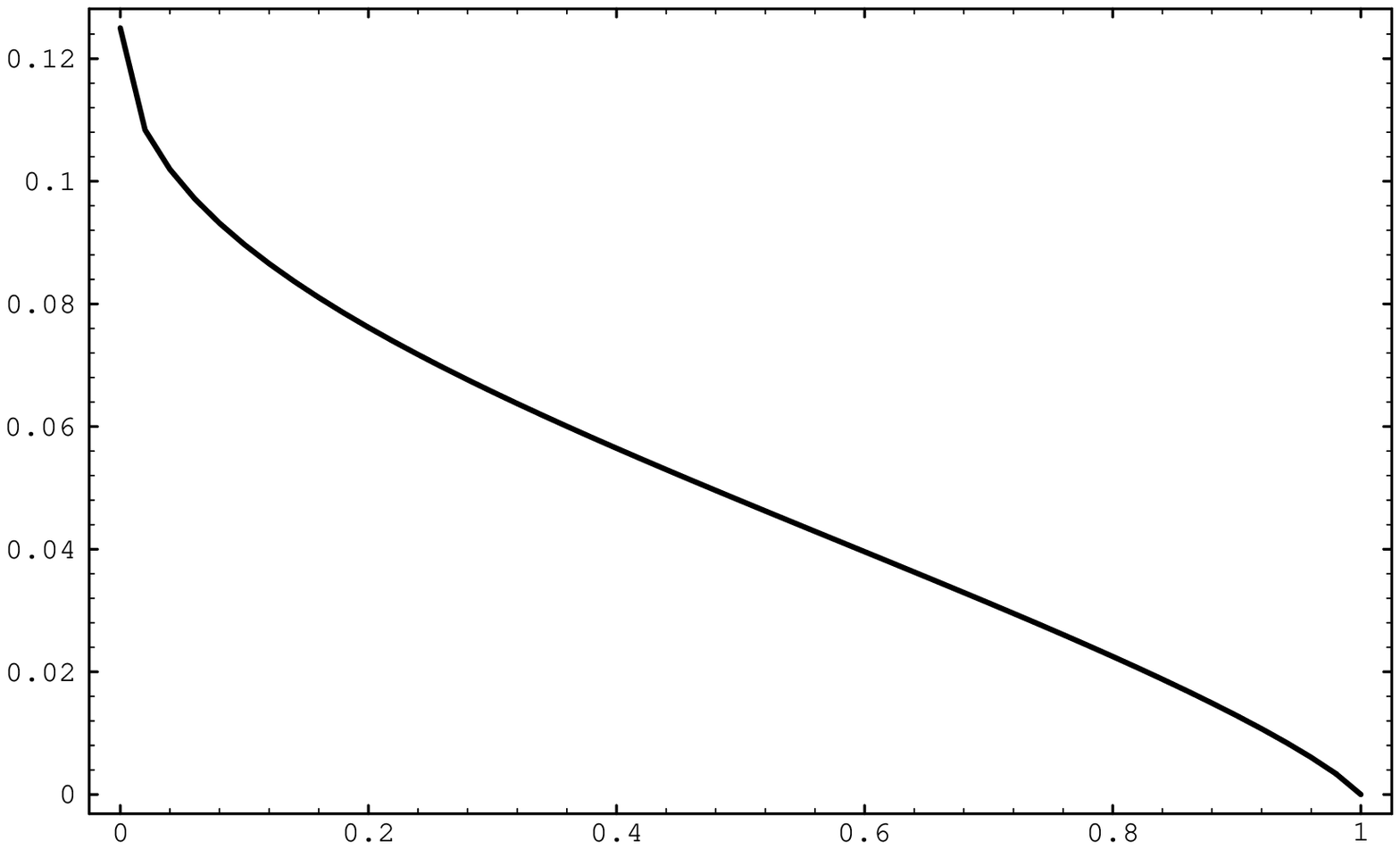,height=25cm,width=15cm}

\vspace{-3cm}

\begin{center}
\hspace{10mm}   
{\bf \large{Figure 9}}
\end{center}
\end{figure}

\newpage

\pagestyle{empty}

\begin{figure}
%\null\vskip -3cm
\hspace{1cm} 
\psfig{figure=ffmin1.ps,angle=90}

%\begin{center}
%\hspace{10mm}   
%{\bf \large{Figure 8}}
%\end{center}

\end{figure}

\newpage

\pagestyle{empty}

%.\vspace{-3cm}
\begin{figure}
%\null\vskip -3cm
\hspace{1cm} \psfig{figure=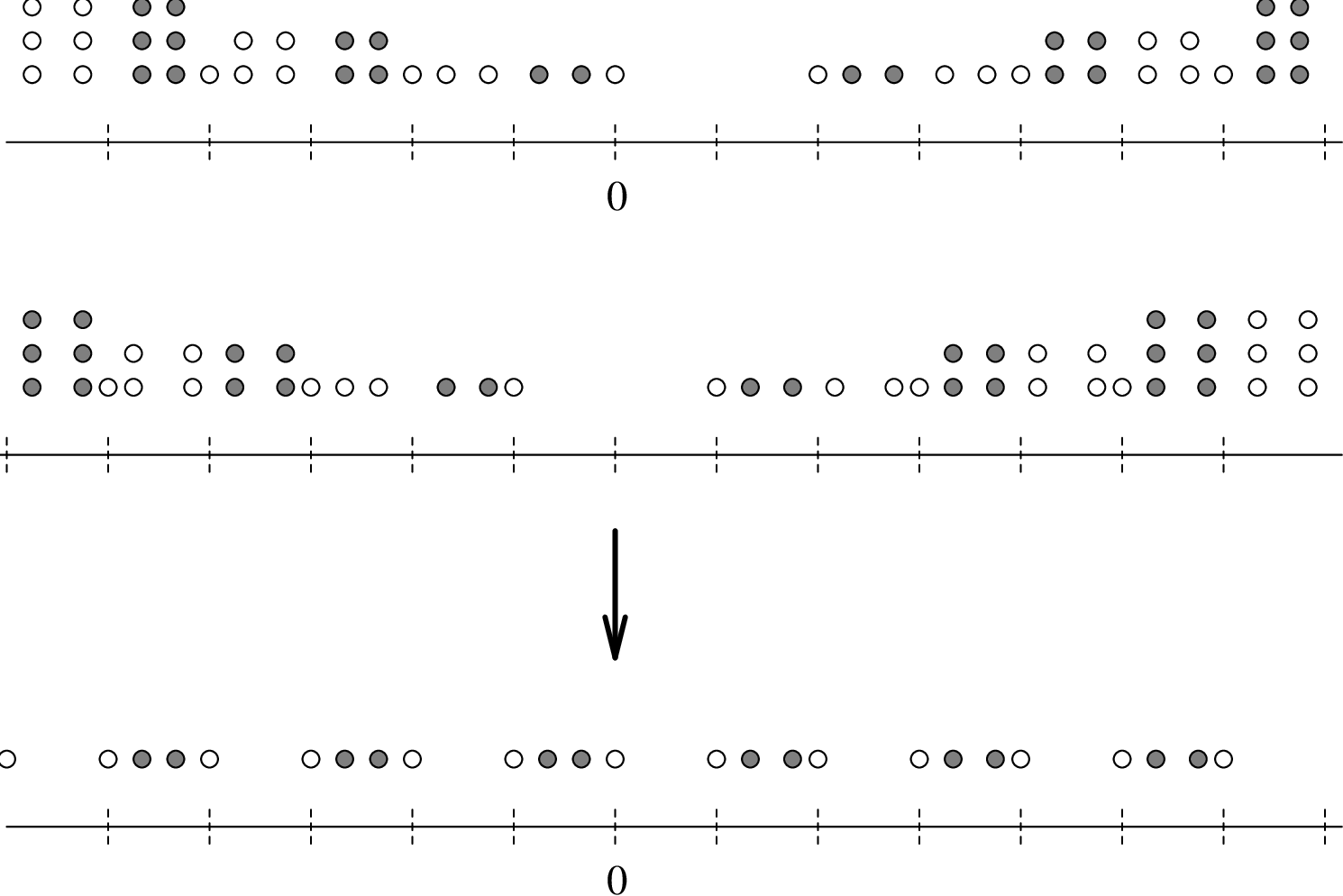,height=15cm,width=15cm}

\vspace{1cm}

\begin{center}
\hspace{10mm}   
{\bf \large{Figure 11}}
\end{center}
\end{figure}


\newpage

\begin{thebibliography}{99}

\bibitem{ISZ} C. Itzykson, H. Saleur and J.B. Zuber, 
{\em Conformal Invariance and Applications to Statistical Mechanics}, 
(World Scientific, Singapore 1988). 
\bibitem{Tsvelik} A.M. Tsvelik, 
{\em Quantum Field Theory in Condensed Matter Physics}, 
Cambridge University Press 1995.  
\bibitem{GMrep} G. Mussardo, {\em Phys. Rep.} {\bf 218} (1992), 215. 
\bibitem{Zam} A.B. Zamolodchikov,
in {\em Advanced Studies in Pure Mathematics} {\bf 19} (1989), 641;
{\it Int. J. Mod. Phys.}{\bf A3} (1988), 743.
\bibitem{roaming} Al.B. Zamolodchikov, Resonance Factorized 
Scattering and Roaming Trajectories, ENS-LPS-335 (1991). 
\bibitem{Z4Zam} A.B. Zamolodchikov, {\em Comm. Math. Phys.}
{\bf 69} (1979), 165. 
\bibitem{ZZ} A.B. Zamolodchikov and Al.B. Zamolodchikov, {\em Ann.Phys.}
{\bf 120} (1979), 253.
\bibitem{AFZ} A.E. Arinshtein, V.A. Fateev and A.B. Zamolodchikov,
{\em Phys. Lett.} {\bf 87B} (1979), 389.
\bibitem{FMS} A. Fring, G. Mussardo and P. Simonetti, 
{\em Nucl. Phys.} {\bf B 393} (1993), 413. 
\bibitem{KM} A. Koubek and G. Mussardo, 
{\em Phys. Lett.} {\bf B 311} (1993), 193. 
\bibitem{Karowski} B. Berg, M. Karowski and P. Weisz, {\em Phys. Rev.} 
{\bf D19} (1979), 2477; \\
M. Karowski and P. Weisz, {\em Nucl. Phys.} {\bf B 139} (1978), 445. 
\bibitem{Smirnov} F.A. Smirnov, {\em Form Factors in Completly 
Integrable Models of Quantum Field Theory} (World Scientific, Singapore 
1992).  
\bibitem{DM} G. Delfino and G. Mussardo, {\em Nucl. Phys.} {\bf B 455} 
(1995), 724.
\bibitem{cluster1} G. Delfino, P. Simonetti and J.L.
Cardy, {\em Phys. Lett.} {\bf B 387} (1996), 327.
\bibitem{cluster2} C. Acerbi, G. Mussardo and A. Valleriani, 
{\em Journ. Phys.} {\bf A 30} (1997), 2895. 
\bibitem{Koberle} R. K\"oberle and J.A. Swieca, {\em Phys. Lett.} {\bf 86B}
(1979), 209; 
\bibitem{Yurov} V.P. Yurov and Al. B. Zamolodchikov,
{\em Int. J. Mod. Phys.} {\bf A6} (1991), 3419.
\bibitem{CMform} J.L. Cardy and G. Mussardo, {\em Nucl. Phys.} {\bf B340}
(1990), 387.
\bibitem{McCoy} B.M. McCoy, C.A. Tracy and T.T. Wu, 
{\em Phys. Rev.} {\bf B 13} (1976), 316; {\em Journ. Math. Phys.} 
{\bf 1977}, 1058. 
%\bibitem{YLZam} Al.B. Zamolodchikov, {\em Nucl. Phys.} {\bf B348} (1991),
%619.
\bibitem{cth} A.B. Zamolodchikov, {\em JEPT Lett.} {\bf 43} (1986), 730.
\bibitem{Cardycth} J.L. Cardy, {\em Phys. Rev. Lett.} {\bf 60}
(1988), 2709.
\bibitem{TBA} Al.B. Zamolodchikov, {\em Nucl. Phys.} 
{\bf B 342} (1990), 695. 
\bibitem{KlMe} T. Klassen and E. Melzer, {\em Nucl. Phys.} 
{\bf B 350} (1991), 635. 
\bibitem{Gran} I.S. Gradshteyn and I.M. Ryzhik, 
{\em Table of Integrals, Series and Products}, Academic Press (1992) 
\end{thebibliography}
\end{document}